\documentclass[preprintnumbers,superscriptaddress,11pt]{revtex4}

\usepackage{palatino}
\usepackage[latin1]{inputenc}
\usepackage{epsf}
\usepackage{amsmath,amssymb}
\usepackage{latexsym}
\usepackage{calc}
\usepackage{color}
\usepackage{shadow}
\usepackage{epsfig}
\usepackage[pdftex, pdfstartview = {FitH}]{hyperref}
\usepackage{graphicx}

\def\bA{{\bf A}}

\def\dulR{{\underline{\underline{\bf R}}}}
\def\dulr{{\underline{\underline{\bf r}}}}

\begin{document}

\title{Mixed quantum-classical dynamics on the exact time-dependent potential energy surface: A fresh look at non-adiabatic processes}
\author{Federica Agostini}
\affiliation{Max-Planck Institut f\"ur Mikrostrukturphysik, Weinberg 2, D-06120 Halle, Germany}
\author{Ali Abedi}
\affiliation{Max-Planck Institut f\"ur Mikrostrukturphysik, Weinberg 2, D-06120 Halle, Germany}
\author{Yasumitsu Suzuki}
\affiliation{Max-Planck Institut f\"ur Mikrostrukturphysik, Weinberg 2, D-06120 Halle, Germany}
\author{E. K. U. Gross}
\affiliation{Max-Planck Institut f\"ur Mikrostrukturphysik, Weinberg 2, D-06120 Halle, Germany}

\begin{abstract}
The exact nuclear time-dependent potential energy surface arises from the exact decomposition of electronic and nuclear motion,
recently presented in  [A. Abedi, N. T. Maitra, and E. K. U. Gross, Phys. Rev. Lett. 105, 123002 (2010)]. Such time-dependent potential
drives nuclear motion and fully accounts for the coupling to the electronic subsystem. We investigate the features of the potential in the
context of electronic non-adiabatic processes and employ it to study the performance of the classical approximation on nuclear dynamics. We
observe that the potential, after the nuclear wave-packet splits at an avoided crossing, develops dynamical steps connecting different
regions, along the nuclear coordinate, in which it has the same slope as one or the other adiabatic surface. A detailed analysis of these
steps is presented for systems with different non-adiabatic coupling strength. The exact factorization of the electron-nuclear wave-function
is at the basis of the decomposition. In particular, the nuclear part is the true nuclear wave-function, solution of a time-dependent
Schroedinger euqation and leading to the exact many-body density and current density. As a consequence, the Ehrenfest theorem can be
extended to the nuclear subsystem and Hamiltonian, as discussed here with an analytical derivation and numerical results.
\end{abstract}

\maketitle

\section{Introduction}\label{sec: intro}
The Born-Oppenheimer (BO)~\cite{BO}, or adiabatic, treatment of the coupled motion of electrons and nuclei is among the most fundamental
approximations in modern condensed-matter theory and forms the basis of our understanding of dynamical processes in molecules and solids. It
offers a practical way to visualize a molecule or solid as a set of nuclei moving on a single potential energy surface (PES) generated by
the electrons in a given eigenstate. However, it is based on the assumption that the electrons adjust instantaneously to adiabatic changes
of the nuclear positions, and a variety of interesting phenomena in physics, chemistry and biology take place in the regime where this
approximation breaks down. Prominent examples are the process of vision~\cite{cerulloN2010,schultenBJ2009,ishidaJPCB2012},
photo-synthesis~\cite{tapaviczaPCCP2011,flemingN2005}, photo-voltaic processes~\cite{rozziNC2013,silvaNM2013,jailaubekovNM2013},
proton-transfer/hydrogen storage ~\cite{sobolewski, varella, hammes-schiffer, marx} as well as phonon-induced superconductivity.

Non-adiabatic molecular processes are usually explained in terms of BOPESs and transitions between the BO electronic states. In this
context, the solution of the time-dependent Schr\"odinger equation (TDSE) is expanded in the complete system of BO electronic states,
leading to a nuclear wave-packet with contributions on several BOPESs that undergo transitions in the regions of strong non-adiabatic
coupling. This approach provides a formally exact description of the complete system if all the electronic states are taken into account.
However, practical applications are limited to a small number of degrees of freedom. For large systems, the only feasible way of dealing
with non-adiabatic processes is the introduction of classical or semi-classical approximations for the nuclear motion, coupled,
non-adiabatically, to the (quantum mechanical) electrons. Although widely investigated~\cite{pechukas,ehrenfest,TSH,kapral-ciccotti}, the
nature of the force driving the classical nuclei in this mixed quantum-classical treatment has not yet been fully identified.

Recently~\cite{steps}, this problem has been addressed from a novel perspective by referring to the exact representation of the full
molecular wave-function~\cite{AMG,AMG2} as a single product of a purely nuclear wave-function and an electronic factor that parametrically
depends on the nuclear coordinates. In this framework, a TDSE for the nuclear wave-function is derived, where a time-dependent potential
energy surface (TDPES) and a time-dependent vector potential arise as exact concepts and provide the \textit{driving force} for the nuclear
evolution.

The present paper discusses situations where the vector potential can be set to zero by an appropriate choice of gauge, thus leaving the
TDPES as the only potential responsible for the nuclear dynamics. In this case, the force on the nuclei, in a classical sense, can be
obtained as the gradient of the TDPES. But, is this the true classical force on the nuclei? We will try to address this issue by employing
the exact TDPES, that is known for the simple system studied here, for the propagation of classical trajectories in order to (i) examine the
quality of the classical approximation for the nuclear motion and (ii) get insight into the properties of approximated classical forces for
an eventual mixed quantum-classical treatment of non-adiabatic processes. Moreover, we will discuss the connections~\cite{steps} between
such novel approach, based on a \textit{single} TDPES, and the well-established description in terms of several static (coupled) BOPESs.

The paper is organized as follows. In Section~\ref{sec: background}, the exact factorization of the time-dependent electron-nuclear
wave-function is presented and the equations that govern the evolution of the electronic and nuclear subsystems are discussed. The TDPES is
investigated and analyzed in detail in Section~\ref{sec: pes} for systems showing different degree of non-adiabaticity. Section~\ref{sec:
dynamics} presents some results obtained by performing classical dynamics on the exact surface and in Section~\ref{sec: ehrenfest} we
discuss the Ehrenfest theorem in the exact factorization representation of the full wave-function. In Section~\ref{sec: conclusion} some
concluding words are given.

\section{Exact decomposition of the electronic and nuclear motion}\label{sec: background}
In the absence of a time-dependent external field, a system of interacting electrons and nuclei is described, 
non-relativistically, by the Hamiltonian
\begin{equation}\label{eqn: hamiltonian}
 \hat H = \hat T_n+\hat H_{BO},
\end{equation}
where $\hat T_n$ is the nuclear kinetic energy operator and 
\begin{equation}\label{eqn: boe}
\hat{H}_{BO}(\dulr,\dulR) = \hat{T}_e(\dulr) + \hat{W}_{ee}(\dulr) + \hat{V}_{en}(\dulr,\dulR) + \hat{W}_{nn}(\dulR),
\end{equation}
is the standard BO electronic Hamiltonian. The symbols $\dulr$ and $\dulR$ are used to collectively indicate the 
coordinates of $N_{e}$ electrons and $N_n$ nuclei, respectively. It has been proved in~\cite{AMG,AMG2} that the
full time-dependent electron-nuclear wave function, $\Psi(\dulr,\dulR,t)$, that is the solution of the TDSE,
\begin{equation}\label{eqn: tdse}
 \hat H\Psi(\dulr,\dulR,t)=i\hbar\partial_t \Psi(\dulr,\dulR,t),
\end{equation}
can be exactly factorized to the correlated product,
\begin{equation}\label{eqn: factorization}
 \Psi(\dulr,\dulR,t)=\chi(\dulR,t)\Phi_\dulR(\dulr,t),
\end{equation}
of the nuclear wave-function, $\chi(\dulR,t)$, and the electronic wave-function, $\Phi_\dulR(\dulr,t)$, that 
parametrically depends on the nuclear configuration and satisfies the partial normalization condition (PNC), 
\begin{equation}
 \int d\dulr \left|\Phi_\dulR(\dulr,t)\right|^2 = 1, \quad\forall\,\,\dulR,t.
\end{equation}
The PNC is an essential element of this representation. Without imposing the PNC, the full wave-function can be 
factorized in many different (unphysical) ways. It is the PNC that makes the factorization~(\ref{eqn:
factorization}) unique up to within a $(\dulR,t)$-dependent gauge transformation, 
\begin{equation}\label{eqn:
gauge}
 \begin{array}{rcl}
  \chi(\dulR,t)\rightarrow\tilde\chi(\dulR,t)&=&e^{-\frac{i}{\hbar}\theta(\dulR,t)}\chi(\dulR,t) \\
  \Phi_\dulR(\dulr,t)\rightarrow\tilde\Phi_\dulR(\dulr,t)&=&e^{\frac{i}{\hbar}\theta(\dulR,t)}\Phi_\dulR(\dulr,t).
 \end{array}
\end{equation}
Another important implication of imposing the PNC is that the diagonal of the $N$-body nuclear density matrix of 
the complete system is equal to $|\chi(\dulR,t)|^2$.  

The stationary variations~\cite{frenkel} of the quantum mechanical action\footnote{The PNC is inserted in the
calculation of the stationary variations of the quantum mechanical action by means of Lagrange multipliers.} w.r.t.
$\Phi_\dulR(\dulr,t)$ and $\chi(\dulR,t)$ lead to the derivation of the equations of motion
\begin{eqnarray}
 \left(\hat{H}_{BO}(\dulr,\dulR)+\hat U_{en}^{coup}[\Phi_\dulR,\chi]-\epsilon(\dulR,t)\right)
 \Phi_{\dulR}(\dulr,t)&=&i\hbar\partial_t \Phi_{\dulR}(\dulr,t)\label{eqn: exact electronic eqn} \\ 
 \left(\sum_{\nu=1}^{N_n} \frac{\left[-i\hbar\nabla_\nu+\bA_\nu(\dulR,t)\right]^2}{2M_\nu} + \epsilon(\dulR,t)
 \right)\chi(\dulR,t)&=&i\hbar\partial_t \chi(\dulR,t). \label{eqn: exact nuclear eqn}
\end{eqnarray}
Here, $\epsilon(\dulR,t)$ is the TDPES,  defined as
\begin{equation}\label{eqn: tdpes}
 \epsilon(\dulR,t)=\left\langle\Phi_\dulR(t)\right|\hat{H}_{BO}+\hat U_{en}^{coup}-i\hbar\partial_t\left|
 \Phi_\dulR(t)\right\rangle_\dulr,
\end{equation}
$\hat U_{en}^{coup}[\Phi_\dulR,\chi]$ is what we name ``electron-nuclear coupling operator'', defined as
\begin{align}
 \hat U_{en}^{coup}[\Phi_\dulR,\chi]=&\sum_{\nu=1}^{N_n}\frac{1}{M_\nu}\left[
 \frac{\left[-i\hbar\nabla_\nu-\bA_\nu(\dulR,t)\right]^2}{2}\right.\label{eqn: enco} \\
 &\left.+\left(\frac{-i\hbar\nabla_\nu\chi}{\chi}+\bA_\nu(\dulR,t)\right)
 \left(-i\hbar\nabla_\nu-\bA_{\nu}(\dulR,t)\right)\right],\nonumber
\end{align}
and $\bA_{\nu}\left(\dulR,t\right)$ is the time-dependent vector potential potential,
\begin{equation}\label{eqn: vector potential}
 \bA_{\nu}\left(\dulR,t\right) = \left\langle\Phi_\dulR(t)\right|-i\hbar\nabla_\nu\left.\Phi_\dulR(t)
 \right\rangle_\dulr.
\end{equation}
The symbol $\left\langle\,\,\cdot\,\,\right\rangle_\dulr$ indicates an integration over electronic 
coordinates only. 

In Eqs.~(\ref{eqn: exact electronic eqn}) and~(\ref{eqn: exact nuclear eqn}), $\hat
U_{en}^{coup}[\Phi_\dulR,\chi]$, $\epsilon(\dulR,t)$ and $\bA_{\nu}\left(\dulR,t\right)$ mediate the coupling
between the electronic and nuclear motions in a formally exact way. The electron-nuclear coupling operator, $\hat
U_{en}^{coup}[\Phi_\dulR,\chi]$, in the electronic equation~(\ref{eqn: exact electronic eqn}),  
depends on the nuclear wave-function and the first and second derivatives of the electronic wave-function with 
respect to the nuclear coordinates. This operator includes the coupling to the nuclear subsystem beyond the parametric dependence in the BO
Hamiltonian $\hat H_{BO}(\dulr,\dulR)$. The nuclear equation~(\ref{eqn: exact nuclear eqn}), on the other hand, has a particularly
appealing form of a Schr\"odinger equation that contains a time-dependent vector potential~(\ref{eqn: vector potential}) and a
time-dependent scalar potential~(\ref{eqn: tdpes}) that uniquely~\footnote{The scalar and vector potentials are uniquely determined up to
within a gauge transformation, given in Eqs.~(\ref{eqn: transformation of epsilon}) and~(\ref{eqn: transformation of A}). However, as
expected, the nuclear Hamiltonian in Eq.~(\ref{eqn: exact nuclear eqn}) is form-invariant under such transformations.}
govern the nuclear dynamics and yield the nuclear wave-function. $\chi(\dulR,t)$ is interpreted as the nuclear wave-function since it leads
to an $N$-body nuclear density, $\Gamma(\dulR,t)=\vert\chi(\dulR,t)\vert^2$, and an $N$-body current density, 
${\bf J}_\nu(\dulR,t)=Im(\chi^*\nabla_\nu\chi)+ \Gamma(\dulR,t){\bf A}_\nu$,  which reproduce the true nuclear $N$-body
density and current density obtained from the full wave-function $\Psi(\dulr,\dulR,t)$~\cite{AMG2}. The uniqueness of $\epsilon(\dulR,t)$
and $\bA_{\nu}(\dulR,t)$ can be straightforwardly proved by following the steps of the current-density version~\cite{Ghosh-Dhara} of the
Runge-Gross theorem~\cite{RGT}. The scalar potential and the vector potential transform as  
\begin{eqnarray}
\tilde{\epsilon}(\dulR,t) &=& \epsilon(\dulR,t)+\partial_t\theta(\dulR,t)\label{eqn: transformation of epsilon} \\
 \tilde{\bf A}_{\nu}(\dulR,t) &=& {\bf A}_{\nu}(\dulR,t)+\nabla_\nu\theta(\dulR,t),\label{eqn: transformation of A}
\end{eqnarray}
under the gauge transformation~(\ref{eqn: gauge}).

\section{Time-dependent potential energy surface}\label{sec: pes}
In this work, we present a detailed study of the TDPES for strongly coupled electronic and nuclear motions. In
order to obtain the TDPES, the full electron-nuclear wave-function has to be calculated. Therefore, we need to
choose a system that is simple enough to allow for a numerically exact treatment and that nevertheless exhibits
characteristic features associated with non-adiabatic dynamics. Here, we employ the model of Shin and
Metiu~\cite{MM}, consisting of three ions and a single electron, as depicted in Fig.~\ref{fig: metiu
model}.
\begin{figure}[h!]
 \centering
 \includegraphics{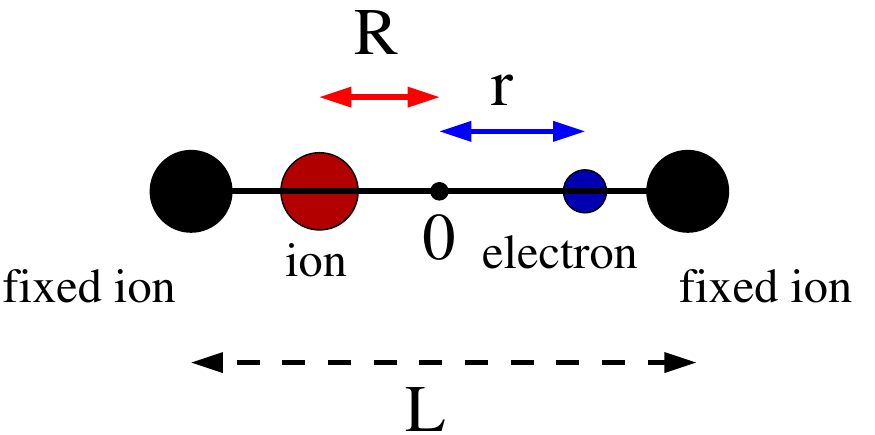}
 \caption{Schematic representation of the model system described by the 
Hamiltonian~(\ref{eqn: metiu-hamiltonian}). $R$ and $r$ indicate the coordinates of the moving ion and electron,
respectively, in one dimension. $L$ is the distance between the fixed ions.}
 \label{fig: metiu model}
\end{figure}
Two ions are fixed at a distance of $L=19.0$~$a_0$, the third ion and the electron are free to move in one 
dimension along the line joining the two fixed ions. The Hamiltonian of this system reads
\begin{align}\label{eqn: metiu-hamiltonian}
 \hat{H}(r,R)= &-\frac{1}{2}\frac{\partial^2}{\partial r^2}-\frac{1}{2M}\frac{\partial^2}{\partial R^2}  +
 \frac{1}{\left|\frac{L}{2}
 -R\right|}+\frac{1}{\left|\frac{L}{2} + R\right|}-\frac{\mathrm{erf}\left(\frac{\left|R-r\right|}{R_f}\right)}
 {\left|R - r\right|}\nonumber\\
 &-\frac{\mathrm{erf}\left(\frac{\left|r-\frac{L}{2}
 \right|}{R_r}\right)}{\left|r-\frac{L}{2}\right|}-\frac{\mathrm{erf}\left(\frac{\left|r+\frac{L}{2}\right|}
 {R_l}\right)}{\left|r+\frac{L}{2}\right|}.
\end{align}
Here, the symbols $\dulr$ and $\dulR$ are replaced by $r$ and $R$, the coordinates of the electron and the 
movable ion measured from the center of the two fixed ions and $M=1836$ is the mass of the movable ion. The 
parameters $R_f$, $R_l$ and $R_r$ specify the interactions between the charged particles and can be tuned to have
different couplings between the electronic and nuclear motions. 

To obtain the TDPES, we first solve the TDSE~(\ref{eqn: tdse}) for the complete system and obtain the full 
wave-function, $\Psi(r,R,t)$. This is done by the numerical integration of the TDSE using SPO-technique~\cite{spo},
with the time-steps of $2.4\times10^{-3}$~$fs$ (or $0.1$~$a.u.$). The nuclear density is calculated, at each time, as the
marginal probability of the configuration $\dulR$~\footnote{We reintroduce the bold-double underlined symbols for electronic and nuclear
positions whenever the statements have general validity. } from the full wave-function
\begin{equation}
 \left|\chi(\dulR,t)\right|^2=\int d\dulr \left|\Psi(\dulr,\dulR,t)\right|^2.
\end{equation}
The phase $S(\dulR,t)$ of $\chi(\dulR,t)$ is determined by the choice of the gauge. We use the exact equality
\begin{equation}\label{eqn: vector-exact}
 \bA_\nu(\dulR,t) = \left|\chi(\dulR,t)\right|^{-2}\mbox{Im}\int d\dulr\,\Psi^*(\dulr,\dulR,t)\nabla_\nu
 \Psi(\dulr,\dulR,t) - \nabla_\nu S(\dulR,t)
\end{equation}
which follows immediately from the definition~(\ref{eqn: vector potential}) of the vector potential by inserting the
factorization~(\ref{eqn: factorization}). The gauge is chosen by setting the vector potential to zero $A(R,t)\equiv0$ in Eq.~(\ref{eqn:
vector-exact}), which is possible in our specific example because we are dealing with a one-dimensional system. Obviously, the
choice of the gauge does not affect any physical observable. $S(R,t)$ is thus determined from the expression
\begin{equation}
 S(R,t)=\int^R dR'
 \left|\chi(R',t)\right|^{-2}\mbox{Im}\int dr\,\Psi^*(r,R',t)\nabla_{R'}\Psi(r,R',t).
\end{equation}
From the calculated exact nuclear wave-function $\chi(\dulR,t)=e^{-\frac{i}{\hbar}S(\dulR,t)}|\chi(\dulR,t)|$, we
obtain the TDPES $\epsilon(\dulR,t)$ from Eq.~(\ref{eqn: tdpes}) by explicitly calculating the electronic
wave-function $\Phi_\dulR(\dulr,t)=\Psi(\dulr,\dulR,t)/\chi(\dulR,t)$. Alternatively, we may invert the nuclear equation~(\ref{eqn: exact
nuclear eqn}). In the gauge we have implemented to perform the calculations, the TDPES alone determines the time evolution of
$\chi(\dulR,t)$. In order to investigate the TDPES in detail, we study its gauge-invariant (GI) and gauge-dependent
(GD) constituents separately (it can be easily proven that
$\tilde{\epsilon}_{GI}(\dulR,t)=\epsilon_{GI}(\dulR,t)$ and
$\tilde{\epsilon}_{GD}(\dulR,t)=\epsilon_{GD}(\dulR,t)+\partial_t\theta(\dulR,t)$ under the transformations in
Eqs.~(\ref{eqn: gauge})), 
\begin{equation}
 \epsilon(\dulR,t)=\epsilon_{GI}(\dulR,t)+\epsilon_{GD}(\dulR,t),
\end{equation}
where
\begin{equation}\label{eqn: gi tdpes}
 \epsilon_{GI}(\dulR,t)=\left\langle\Phi_\dulR(t)\right|\hat{H}_{BO}\left|\Phi_\dulR(t)\right\rangle_\dulr
 +\sum_{\nu=1}^{N_n}\bigg(\frac{\hbar^2}{2M_\nu}\left\langle\nabla_\nu\Phi_\dulR(t)|\nabla_\nu\Phi_\dulR(t)
 \right\rangle_\dulr-\frac{\bA^2_\nu(\dulR,t)}{2M_\nu}\bigg),
\end{equation}
with the second term on the RHS obtained from the action of the electron-nuclear coupling operator in Eq.~(\ref{eqn: enco}) 
on the electronic wave-function, and
\begin{equation}\label{eqn: gd tdpes}
 \epsilon_{GD}(\dulR,t)=\left\langle\Phi_\dulR(t)\right|-i\hbar\partial_t\left|\Phi_\dulR(t)\right\rangle_\dulr.
\end{equation}
The GI part of the TDPES, $\epsilon_{GI}$, is not affected by the gauge transformation~(\ref{eqn: gauge}). The 
GD part, on the other hand, depends on the choice of the gauge. They both have important features~\cite{steps} that will be discussed and
analyzed in the following section.  For this analysis, we will use a representation in terms of the BO electronic states,
$\varphi_{\dulR}^{(l)}(\dulr)$, and BOPESs, $\epsilon_{BO}^{(l)}(\dulR)$, which are the eigenstates and eigenvalues
of the BO electronic Hamiltonian~(\ref{eqn: boe}), respectively. If the full wave-function is expanded in this
basis,
\begin{equation}\label{eqn: expansion of Psi}
 \Psi(\dulr,\dulR,t)=\sum_l F_l(\dulR,t)\varphi_\dulR^{(l)}(\dulr),
\end{equation} 
then the nuclear density may be written as  
\begin{equation}\label{eqn: chi and Fl}
 \left|\chi(\dulR,t)\right| = \sqrt{\sum_{l}\left|F_l(\dulR,t)\right|^2}.
\end{equation}
This identity is obtained by integrating the squared modulus of Eq.~(\ref{eqn: expansion of Psi}) over the electronic 
coordinates with normalized adiabatic states. The exact electronic wave-function may also be expanded in terms of
the BO states,
\begin{equation}\label{eqn: expansion of Phi}
 \Phi_\dulR(\dulr,t)=\sum_l C_l(\dulR,t)\varphi_\dulR^{(l)}(\dulr).
\end{equation} 
The expansion coefficients of Eqs.~(\ref{eqn: expansion of Psi}) and~(\ref{eqn: expansion of Phi}) are related,
\begin{equation}\label{eqn: relation coefficients}
 F_l(\dulR,t)= C_l(\dulR,t)\chi(\dulR,t),
\end{equation}
by virtue of the factorization~(\ref{eqn: factorization}). The PNC then reads
\begin{equation}\label{eqn: PNC on BO}
 \sum_l\left|C_l(\dulR,t)\right|^2=1,\quad\forall\,\,\dulR,t.
\end{equation}
In the cases studied in the following sections, the initial wave-function is the product of a real-valued normalized Gaussian wave-packet,
centered at $R_c=-4.0$~$a_0$ with variance $\sigma=1/\sqrt{2.85}$~$a_0$ (black line in Fig.~\ref{fig: BO-data}), and the second BO
electronic state, $\varphi_{R}^{(2)}(r)$.

\subsection{Steps in the TDPES in strong non-adiabatic regime}
\label{sec: strong coupling}
\begin{figure}[h!]
 \begin{center}
 \includegraphics{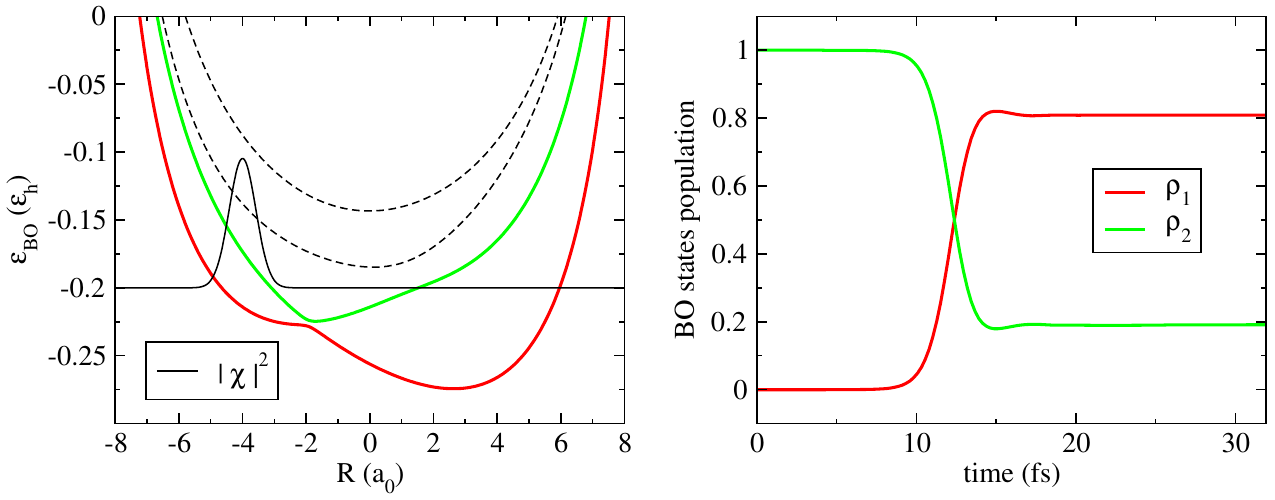}
 \end{center}
 \caption{Left: lowest four BO surfaces, as functions of the nuclear coordinate. The first (red line) and second 
(green line) surfaces will be considered in the actual calculations that follow, the third and forth (dashed black
lines) are shown for reference. The squared modulus (reduced by ten times and rigidly shifted in order to
superimpose it on the energy curves) of the initial nuclear wave-packet is also shown (black line). Right:
populations of the BO states along the time evolution. The strong non-adiabatic nature of the model is
underlined by the population exchange at the crossing of the coupling region.}
 \label{fig: BO-data}
\end{figure}

We first study a case in which the electronic and nuclear motions are strongly coupled. In order to produce that situation, we choose the
parameters of the Hamiltonian~(\ref{eqn: metiu-hamiltonian}) as $R_f=5.0$~$a_0$, $R_l=3.1$~$a_0$ and $R_r=4.0$~$a_0$ such that the first
BOPES, $\epsilon^{(1)}_{BO}$, is strongly coupled to the second BOPES, $\epsilon^{(2)}_{BO}$, around the avoided crossing at
$R_{ac}=-1.90~a_0$ and there is a weak coupling to the rest of the surfaces. The four lowest BOPESs for this set of parameters are shown in
Fig.~\ref{fig: BO-data} (left panel), along with the initial nuclear density. Energies are given in atomic (Hartree) units $\epsilon_h$. The
same figure (right panel) presents the time-evolution of the populations of the BO states,  
\begin{equation}\label{eqn: population BO}
 \rho_{l}(t) = \int d\dulR \left|F_l(\dulR,t)\right|^2,
\end{equation}
and underlines the strong non-adiabatic character of the system with the intense population exchange taking 
place at the passage through the avoided crossing ($t\simeq12$~$fs$).

\begin{figure}[h!]
 \centering
 \includegraphics{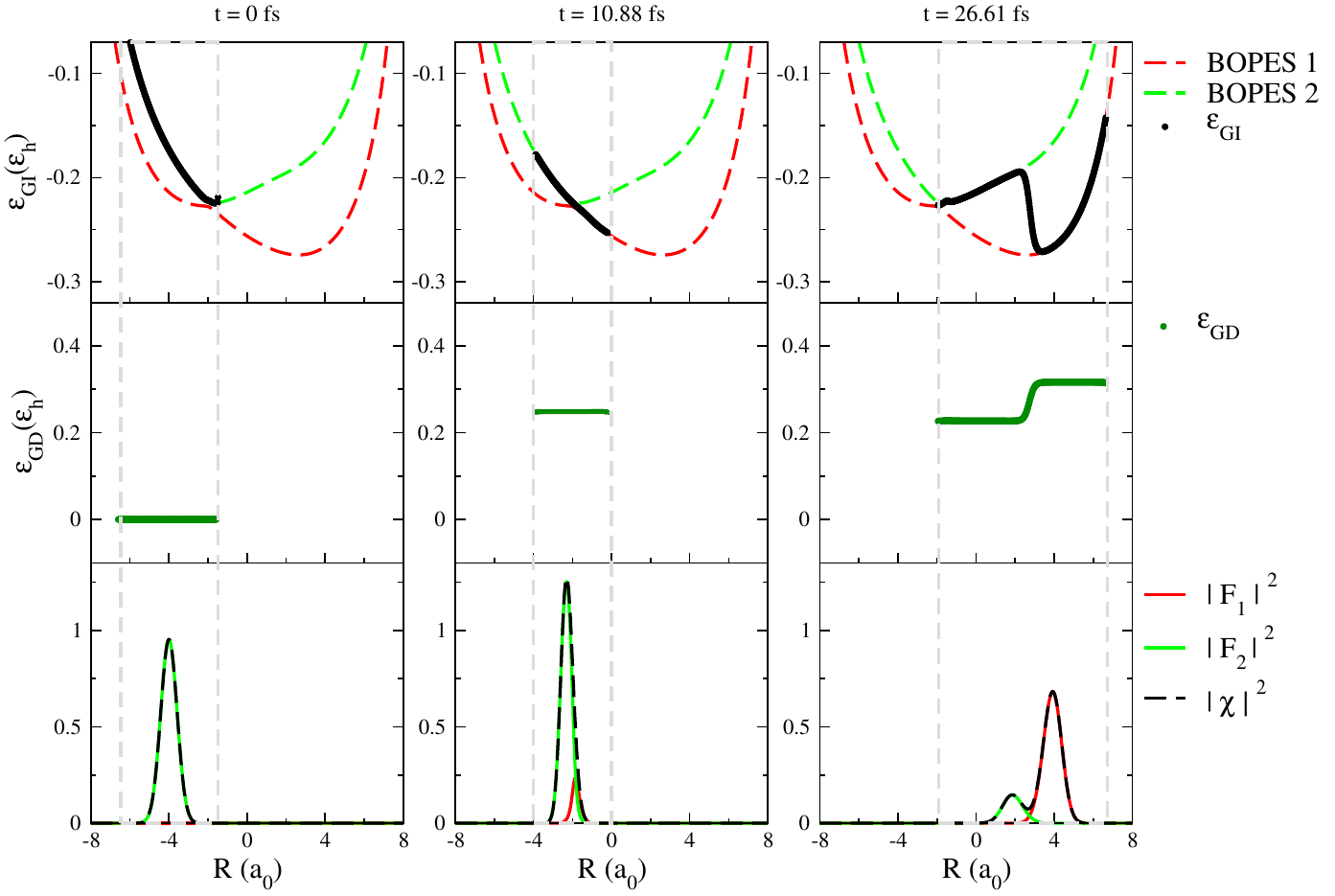}
 \caption{TDPES and nuclear densities at different time-steps, namely $t=0$~$fs$, $t=10.88$~$fs$ and 
$t=26.61$~$fs$. The different panels show: (top) GI part of the TDPES (black dots) and the two lowest BOPESs
(first, dashed red line, and second, dashed green line) as reference; (center) the GD part of the TDPES (green
dots); (bottom) nuclear density (dashed black line) and $|F_l(R,t)|^2$ ($l=1$ red line and $l=2$ green line). The
gray boxes define the regions in $R$-space where the energies have been calculated, since the nuclear density is
(numerically) not zero.}
 \label{fig: snapshots strong}
\end{figure}

As recently discussed~\cite{steps}, the GI part of the TDPES~(\ref{eqn: gi tdpes}) shows, in general, two distinct
features: (i) in the vicinity of the avoided crossing, as the nuclear wave-packet passes through the region of
non-adiabatic coupling between different BOPESs, $\epsilon_{GI}(R,t)$ resembles the \textit{diabatic} surface that
smoothly connects the two adiabatic surfaces; (ii) a bit further away from the avoided crossing, it shows \textit{dynamical
steps} between regions in $R$-space where it is on top of one or the other BOPES. The GD part of the
TDPES~(\ref{eqn: gd tdpes}), on the other hand, is a piecewise constant function of the nuclear coordinate. This
is illustrated in detail in Fig.~\ref{fig: snapshots strong} that contains the GI part of the TDPES (upper
panel), the GD part of the TDPES (middle panel) and the nuclear density together with $|F_1|^2$ and $|F_2|^2$
(lower panel) for three different snapshots of time. In all the plots, the regions highlighted within the boxes
are the regions which we refer to in the following discussion. Outside such regions, the value of the nuclear
density drops under the numerical accuracy and the resulting potentials are not meaningful. That is why the TDPES
are trimmed. The left panels show, at the initial time-step, (top) the GI part of the TDPES (black dots), with the
two lowest BOPESs ($\epsilon_{BO}^{(1)}(R)$, dashed red line, and $\epsilon_{BO}^{(2)}(R)$, dashed green line) as
reference, (center) the GD part of the exact potential (dark-green dots) and (bottom) the nuclear density (dashed
black line) and its components from on the BO states (see Eq.~(\ref{eqn: chi and Fl})), $|F_1(R,t)|^2$ (red line)
and $|F_2(R,t)|^2$ (green line).
At time $t=0$~$fs$, the electronic wave-function, $\Phi_R(r,t)$, coincides with the second adiabatic state 
$\varphi_R^{(2)}(r)$, therefore the GI component of the TDPES is identical with $\epsilon_{BO}^{(2)}(R)$, apart
from a slight deviation due to the second term in Eq.~(\ref{eqn: gi tdpes}). This is easily confirmed by the
expression of $\epsilon_{GI}(R,t)$ in terms of the BO states and energies
\begin{align}
 \epsilon_{GI}(R,t)&=\sum_{l}\left|C_l(R,t)\right|^2\epsilon_{BO}^{(l)}(R)+\frac{\hbar^2}{2M}
 \left[\sum_{l,k}C_l^*(R,t)C_k(R,t)d_{lk}^{(2)}(R)\right.
 \label{eqn: gi tdpes on BO}\\
 &\left.\sum_{l,k}\left({C_l^*}'(R,t)C_k(R,t)-C_l^*(R,t)C_k'(R,t)\right)d_{lk}^{(1)}(R)+ 
 \sum_l\left|C_l'(R,t)\right|^2\right],\nonumber
\end{align}
where we use the prime to indicate the spatial derivative of the coefficients and we introduced the 
non-adiabatic couplings
\begin{eqnarray}
 d_{lk}^{(1)}(R)=&\left\langle\varphi_{R}^{(l)}\right.\left|\nabla_{R}\varphi_{R}^{(k)}\right\rangle_r &
 =-{d_{kl}^{(1)}}^*(R) \\
 d_{lk}^{(2)}(R)=&\left\langle\nabla_{R}\varphi_{R}^{(l)}\right.\left|\nabla_{R}\varphi_{R}^{(k)}\right\rangle_r &
 = {d_{kl}^{(2)}}^*(R).
\end{eqnarray}
The leading term in Eq.~(\ref{eqn: gi tdpes on BO}) is the average of the BOPESs weighted by $\left|C_l(R,t)
\right|^2$, since the second term is $\mathcal O(M^{-1})$. The GD component of the TDPES in Eq.~(\ref{eqn: gd
tdpes}), in terms of the BO states, becomes 
\begin{equation}\label{eqn: gd tdpes on BO}
 \epsilon_{GD}(R,t) = \sum_{l}\left|C_l(R,t)\right|^2\dot\gamma_{l}(R,t)
\end{equation}
where $\dot\gamma_{l}(R,t)$ is the time-derivative of the phase of the coefficients $C_l(R,t)=
e^{\frac{i}{\hbar}\gamma_l(R,t)}|C_l(R,t)|$. The nuclear density, along with its components on the BO states from
Eq.~(\ref{eqn: chi and Fl}), is presented in the bottom panels of Fig.~\ref{fig: snapshots strong}. At the initial time,
$\left|\chi(R,t)\right|^2=\left|F_2(R,t)\right|^2$.

At $t=10.88$~$fs$ in Fig.~\ref{fig: snapshots strong} (central panels), (top) the GI part of the TDPES resembles 
the diabatic surface~\cite{MM} that smoothly passes through the avoided crossing. This behavior allows the nuclear
density moving on the upper BOPES to be partially ``transferred'' to the lower state, as the consistent increase of
the population of state $\varphi_R^{(1)}(r)$ (red curve in the bottom plot in Fig.~\ref{fig: snapshots strong})
confirms. In the region highlighted by the dashed box, the GD part of the exact potential is constant, therefore,
it does not affect nuclear dynamics.

At later times ($t=26.61$~$fs$ shown in the right panels of Fig.~\ref{fig: snapshots strong}), when the nuclear 
wave-packet has split at the avoided crossing, both components of the TDPES present a pronounced stepwise behavior:
the GI part follows one or the other BOPES in different regions of $R$-space that are connected by a step, whereas
the GD part is stepwise constant, with steps appearing in the same region.

The overall shape of the TDPES, at initial times, is determined by the GI part, as the effect of the GD part is no more than a constant
shift. Hence, the TDPES, that drives the nuclear dynamics, behaves like a diabatic surface and ``opens'' in the direction of the
wave-packet's motion in order to facilitate the population exchange between the adiabatic states. After the wave-packet splits at the
avoided crossing, in different regions in $R$-space, the TDPES is parallel to one or the other BOPES and a step forms in the
transition region. Therefore, the motion of the components $F_l(R,t)$ of the nuclear wave-packet is driven by single adiabatic
surfaces and not (like, e.g., in Ehrenfest dynamics) by an average electronic potential. This feature is reminiscent of the way the
well-known \textsl{trajectory surface hopping} (TSH) scheme~\cite{TSH} deals with the non-adiabatic dynamics. In this approach, the
components (in our case identified by the symbol $|F_l(R,t)|^2$) of the nuclear density on different BO states are represented by
\textsl{bundles} of classical trajectories evolving, independently from one another, on different BO surfaces. The ratio of the total number
of trajectories occupying, at each time, the surfaces approximates the population $\rho_l$ of the corresponding BO state. The success of
this method in reproducing non-adiabatic processes becomes clear in the light of the fact that the exact TDPES itself is parallel to
different BOPESs in different regions along the nuclear coordinate. The usually abrupt transitions between the adiabatic surfaces, i.e.,
the steps in the exact treatment, are reminiscent to the stochastic jumps between BO surfaces in TSH.

\subsection{Analysis of the steps}
\begin{figure}[h!]
 \begin{center}
 \includegraphics{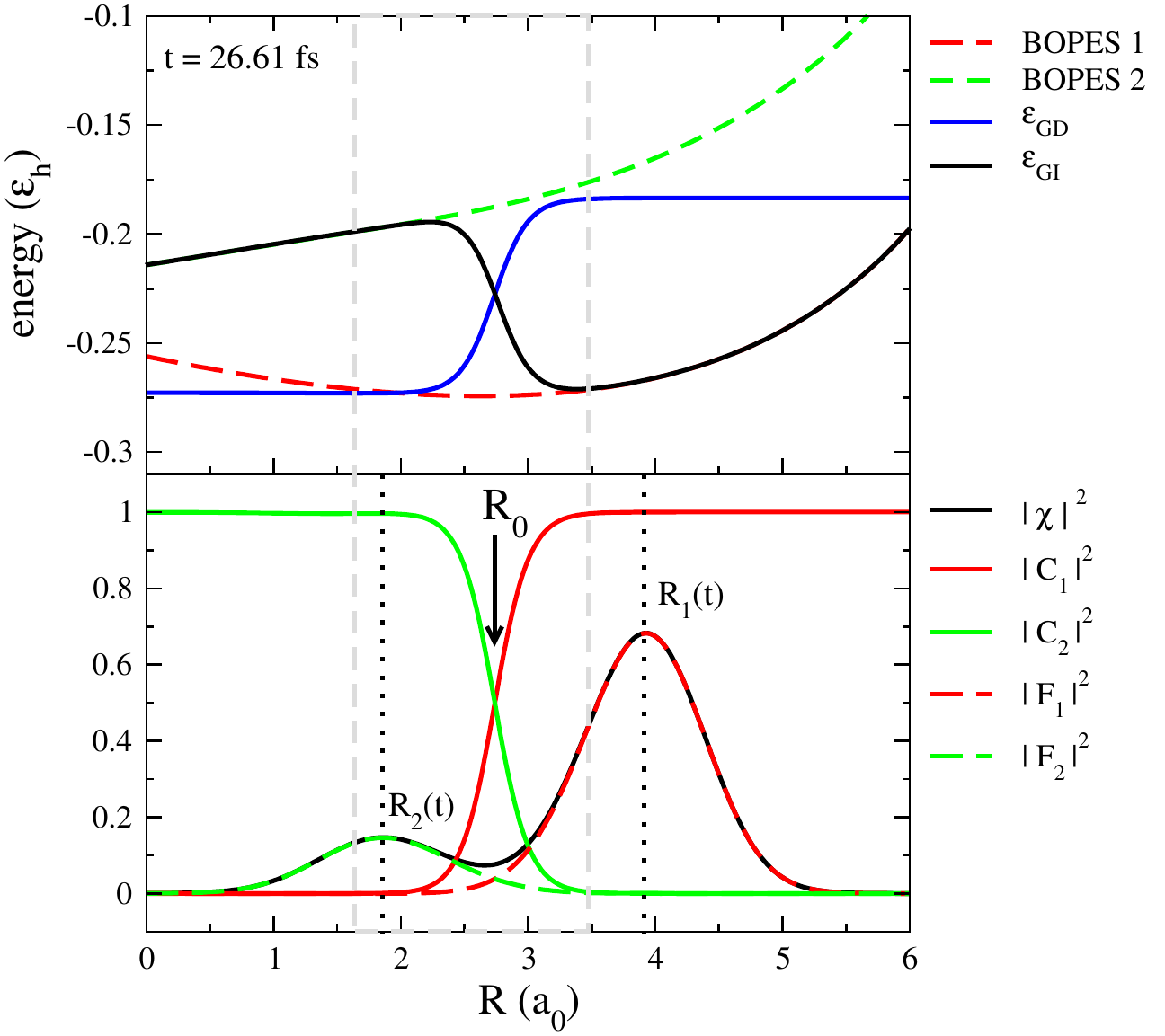}
 \end{center}
 \caption{Top: GI part (black line) and the GD part (blue line, rigidly shifted along the energy axis) of the
exact potential at time $t=26.61$~$fs$. The first (dashed red) and second (dashed green) BOPESs are shown as 
reference. Bottom: coefficients $|F_l(R,t)|^2$ of the expansion of the full wave-function (Eq.~(\ref{eqn: expansion
of Psi})) on the BO states ($l=1$ dashed red line, $l=2$ dashed green line) and coefficients $|C_l(R,t)|^2$ of the
expansion of the electronic wave-function ($l=1$ continuous red line, $l=2$ continuous green line); the black line
represents the nuclear density. $R_0$ is the position where the coefficients $|F_1(R,t)|^2$ and $|F_2(R,t)|^2$ have
the same value and the dashed box highlights the region of the step.}
 \label{fig: steps analysis}
\end{figure}

The behavior of the GI part of the TDPES is mainly determined by the first term in Eq.~(\ref{eqn: gi tdpes on BO}). The steps appear in
the region around $R_0$, the cross-over of $|F_1(R,t))|^2$ and $|F_2(R,t))|^2$. In particular, at this point
$|F_1(R_0,t)|^2=|F_2(R_0,t)|^2=|X(t)|$ and, irrespective of this value, the expansion coefficients in the
electronic wave-function~({\ref{eqn: expansion of Phi}}) have the value $|C_1(R_0,t)|^2=|C_2(R_0,t)|^2=1/2$. This
relation holds as consequence of Eq.~(\ref{eqn: relation coefficients}), which can be written as
\begin{equation}
 \left|C_l(R_0,t)\right|^2 = \frac{\left|F_l(R_0,t)\right|^2}{\left|F_1(R_0,t)\right|^2+\left|F_2(R_0,t)
 \right|^2}=\frac{1}{2}\quad \mbox{with}\quad l=1,2,
\end{equation}
and is clearly shown in Fig.~\ref{fig: steps analysis}.
Here we present, in the upper panel, the GI part (black line) and the GD part (blue line, rigidly shifted along 
the energy axis) of the exact potential at time $t=26.62$~$fs$. The BO surfaces (dashed red and green lines) are
also plotted as reference. In the lower panel, we plot the coefficients of the expansions in Eq.~(\ref{eqn:
expansion of Psi}) (dashed red and green lines) and in Eq.~(\ref{eqn: expansion of Phi}) (continuous red and green
lines). The continuous black line represents the nuclear density.

The expression of the GI component of the TDPES for a two-state system, from Eq.~(\ref{eqn: gi tdpes on BO}), is
\begin{equation}
 \epsilon_{GI}(R,t) \simeq \left|C_1(R,t)\right|^2
\epsilon_{BO}^{(1)}(R)+\left|C_2(R,t)\right|^2\epsilon_{BO}^{(2)}(R),
\end{equation}
neglecting terms $\mathcal O(M^{-1})$. If $\left|C_l(R,t)\right|^2$ is Taylor-expanded around $R_0$, up to within 
the linear deviations,
\begin{eqnarray}
 \left|C_{\mathop{}_2^1}(R,t)\right|^2&\simeq&\left.\frac{\left|F_{\mathop{}_2^1}(R,t)\right|^2}{\left|\chi(R,t)
 \right|^2}\right|_{R_0}+ \left.\nabla_{R}\frac{\left|F_{\mathop{}_2^1}(R,t)\right|^2}{\left|\chi(R,t)\right|^2}
 \right|_{R_0} (R-R_0)\nonumber \\
 &=&\frac{1}{2}\pm \frac{\alpha(t)}{2}\left(R-R_0\right),
\end{eqnarray}
one can identify the parameter $\alpha(t)$, defined as
\begin{equation}\label{eqn: definition of alpha}
 \alpha(t) = \frac{\left(\nabla_R\left|F_1(R,t)\right|\right)_{R_0}-\left(\nabla_R\left|F_2(R,t)
 \right|\right)_{R_0}}{\left|X(t)\right|},
\end{equation}
where $\alpha(t)$ is the slope of the coefficients in the step region from which the width of the region can 
be determined. Using the relation $0\leq\left|C_1(R,t)\right|^2\leq 1$, we get
\begin{equation}
 0\leq \frac{1}{2}+\frac{\alpha(t)}{2}\left(R-R_0\right)\leq 1\quad\mbox{with}\quad\frac{\Delta R}{2}=
 \left|R-R_0\right|\leq \frac{1}{\alpha(t)}.
\end{equation}
Therefore, $\Delta R$ is small because the step is steep, as consequence of a large $\alpha(t)$. $\alpha(t)$ 
can be large either because $|X(t)|$ is small, i.e., the cross-over is located in a region of small nuclear
density, or because the terms in the numerator of Eq.~(\ref{eqn: definition of alpha}) have opposite slopes at
$R_0$ (this is the case depicted in Fig.~\ref{fig: steps analysis}). Outside the region $\Delta R$, one or the
other coefficients $|C_l(R,t)|^2$ dominates, thus leading to 
\begin{equation}\label{eqn: gi tdpes outside step
region}
 \epsilon_{GI}(R,t)=\left\lbrace 
 \begin{array}{cc}
  \epsilon_{BO}^{(2)}(R), & R< R_0 \\
  & \\
  \epsilon_{BO}^{(1)}(R), & R> R_0.
 \end{array}
 \right.
\end{equation}
The GD part of the TDPES can be analyzed similarly: $\epsilon_{GD}(R,t)$ from Eq.~(\ref{eqn: gd tdpes on BO}) may
be written, in terms of the two BO states, as 
\begin{equation}
 \epsilon_{GD}(R,t) = \left|C_1(R,t)\right|^2 \dot\gamma_1(R,t)+\left|C_2(R,t)\right|^2\dot\gamma_2(R,t)
\end{equation}
and we recall that $\gamma_l(R,t)$ is the phase of the coefficient $C_l(R,t)$. As in Eq.~(\ref{eqn: gi tdpes
outside step region}), outside the step region, this part of the potential becomes
\begin{equation}\label{eqn: gd tdpes outside step region}
 \epsilon_{GD}(R,t)=\left\lbrace 
 \begin{array}{cc}
  \dot\gamma_2(R,t), & R< R_0 \\
  & \\
  \dot\gamma_1(R,t), & R> R_0.
 \end{array}
 \right.
\end{equation}
Moreover, Fig.~\ref{fig: steps analysis} shows that in these regions $\dot\gamma_1(R,t)$ and $\dot\gamma_2(R,t)$ 
are constant functions of $R$. This is a consequence of the gauge we chose. The gauge condition,
$A(R,t)=\langle\Phi_R(t)|-i\hbar\nabla_R\Phi_R(t)\rangle_r =0$, in terms of the two BO states involved in the
dynamics, reads
\begin{align}
 0=\sum_{l=1,2}\left|C_l(R,t)\right|^2\nabla_R\gamma_l(R,t)-\frac{i\hbar}{2}\nabla_R\sum_{l=1,2}\left|C_l(R,t)
 \right|^2 \nonumber \\
 -i\hbar\sum_{l,k=1,2}C_l^*(R,t)C_k(R,t)d_{lk}^{(1)}(R).
\end{align}
However, the second term of the RHS is identically zero, due to the PNC in Eq.~(\ref{eqn: PNC on BO}), and the 
third term can be neglected, due to the presence of the non-adiabatic couplings, $d_{lk}^{(1)}(R)$, that are small
far from the avoided crossing. The gauge condition then states
\begin{equation}
 \left|C_1(R,t)\right|^2\nabla_R\gamma_1(R,t) = -\left|C_2(R,t)\right|^2\nabla_R\gamma_2(R,t),
\end{equation}
or equivalently
\begin{eqnarray}
 \nabla_R\gamma_2(R,t) = 0 &\quad\mbox{for}\quad R<R_0\quad&\mbox{where}\quad\left|C_1(R,t)\right|^2=0 \\
 \nabla_R\gamma_1(R,t) = 0 &\quad\mbox{for}\quad R>R_0\quad&\mbox{where}\quad\left|C_2(R,t)\right|^2=0.
\end{eqnarray}
We obtain $\gamma_l(R,t)=\Gamma_l(t)$, namely the phase of the coefficient $C_l(R,t)$ is only a function of time 
(constant in space) in the region where the squared modulus of the corresponding coefficient is equal to unity.
Similarly, $\dot\gamma_l(R,t)=\dot\Gamma_l(t)$, as shown in Fig.~\ref{fig: steps analysis}.

In the step region, around $R_0$, the expression of the TDPES can be approximated as
\begin{align}
 \epsilon(R,t) =& \frac{\epsilon_{BO}^{(1)}(R)+\epsilon_{BO}^{(2)}(R)}{2}+\frac{\dot\gamma_1(R,t)+
 \dot\gamma_2(R,t)}{2}\nonumber \\
 &+\alpha(t)\left[\frac{\epsilon_{BO}^{(1)}(R)-\epsilon_{BO}^{(2)}(R)}{2}+\frac{\dot\gamma_1(R,t)-
 \dot\gamma_2(R,t)}{2}\right](R-R_0). \label{eqn: full potential at R0}
\end{align}
The first two terms on the RHS are the average of the BO energies plus the average value of the time-derivative of
the phases $\gamma_1(R,t)$ and $\gamma_2(R,t)$; the terms in square brackets are the energy gaps between the BO 
surfaces and between the time-derivative of the phases, which give the contribution proportional to the parameter
$\alpha(t)$. From Fig.~\ref{fig: steps analysis}, we notice that, around $R_0$, the slope of $\epsilon_{GD}$ is
opposite to the slope of $\epsilon_{GI}$ and this is a general feature in the studied system (in the absence of a
time-dependent external field). Therefore, the GD part reduces the height of the steps in the GI part.
We will see the effect of this contribution on (classical) nuclear dynamics in the section~\ref{sec: dynamics}.

\subsection{Steps in the TDPES in weak non-adiabatic regime}
\begin{figure}[h!]
 \begin{center}
 \includegraphics{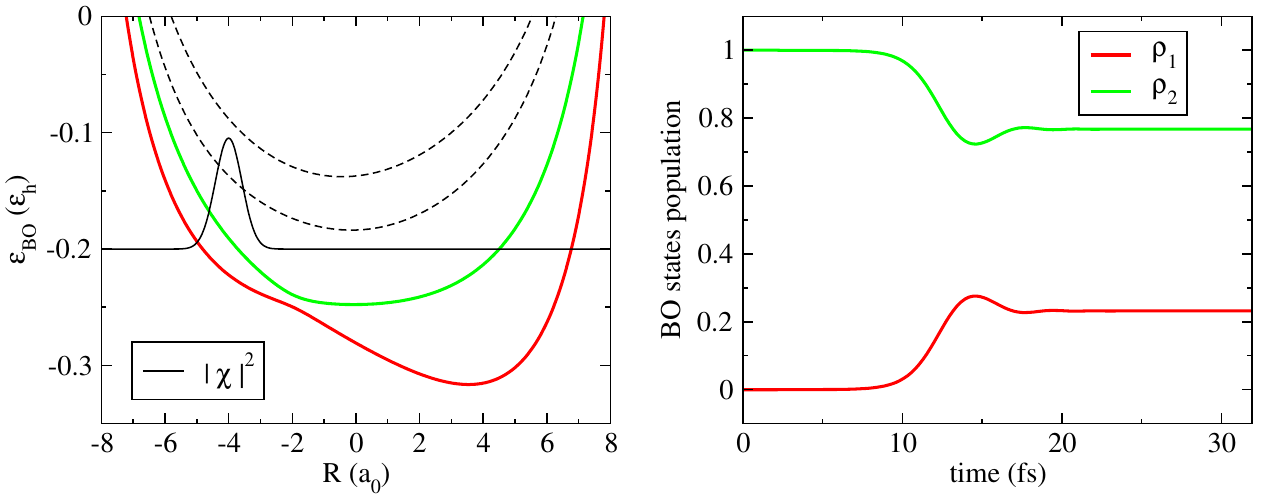}
 \end{center}
 \caption{Same as Fig.~\ref{fig: BO-data} but for weaker non-adiabatic coupling between the two lowest BO 
 states.}
 \label{fig: BO-data weak}
\end{figure}

\begin{figure}[h!]
 \centering
 \includegraphics{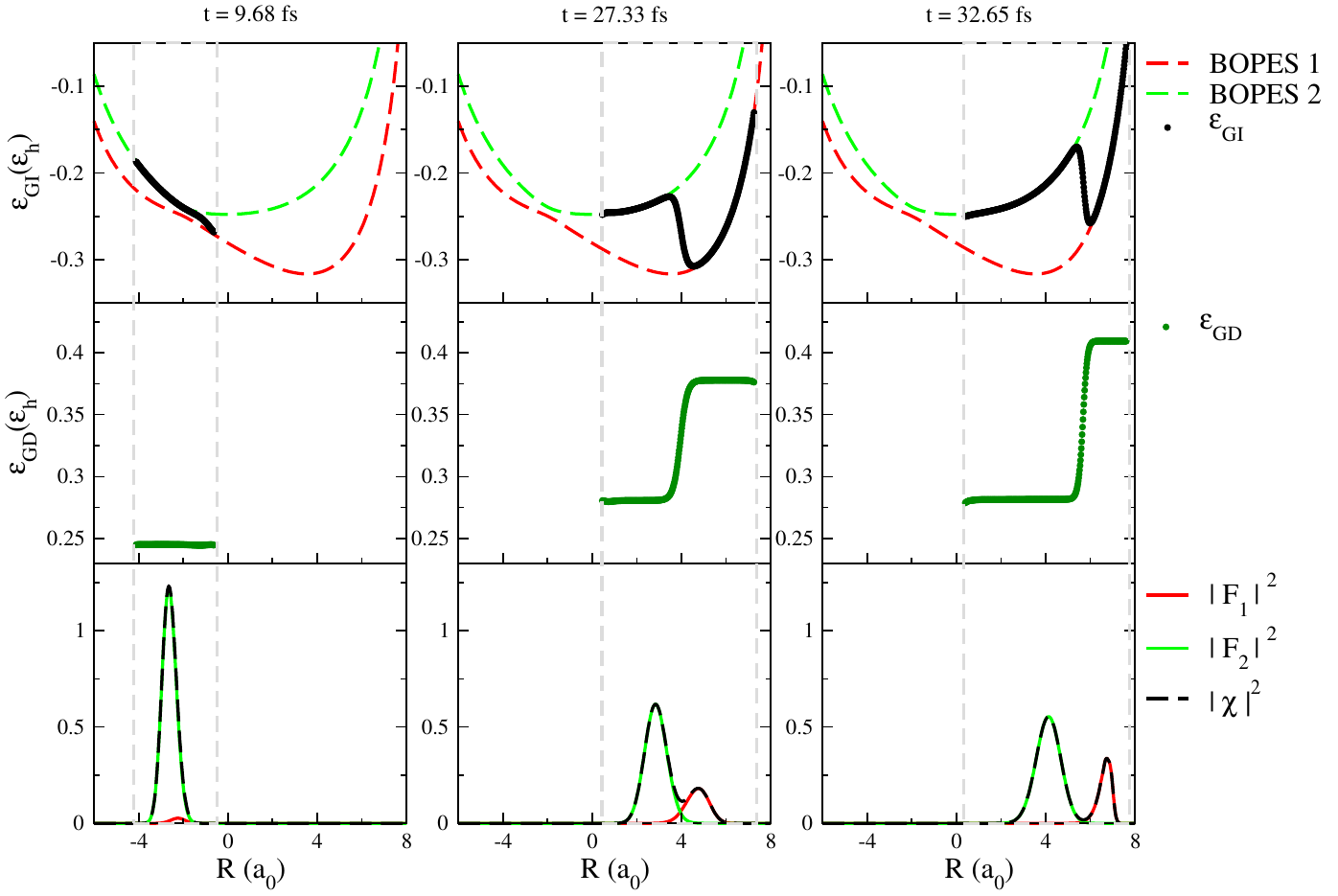}
 \caption{Same as Fig.~\ref{fig: snapshots strong} but for a weaker non-adiabatic coupling between the two 
lowest BO states, at time-steps $9.68$~$fs$, $27.33$~$fs$ and $32.65$~$fs$.}
 \label{fig: snapshots weak}
\end{figure}

In this section, we study a case of weaker non-adiabatic coupling between the two lowest BO states. In order to
make the coupling weaker, we choose the parameters in the Hamiltonian~(\ref{eqn: metiu-hamiltonian}) as
$L=19.0$~$a_0$, $R_f=3.8$~$a_0$, $R_l=2.0$~$a_0$ and $R_r=5.5$~$a_0$. The BO surfaces, along with the evolution of
the populations of the BO states, are shown in Fig.~\ref{fig: BO-data weak}.
The initial conditions for the dynamical evolution of this system are the same as in the previous example, 
however the coupling between the two lowest electronic states is weaker, thus leading to a reduced
population exchange, clearly shown in Fig.~\ref{fig: BO-data weak} (right panel). Nonetheless, the process
described here shows similarities to the previous case, as can be seen
from Fig.~\ref{fig: snapshots weak}.
The GI part of the TDPES presents again two main features, (i) the diabatization at the avoided crossing, 
when the nuclear wave-packet crosses the region of relatively strong non-adiabatic coupling and (ii) the steps at 
the cross-over of $|F_1(R,t)|^2$ and $|F_2(R,t)|^2$, signature of the splitting of the nuclear density. The GD part
is either constant, before the splitting at the avoided crossing, or stepwise constant, with
steps appearing in the same region as the steps in the GI term, but with opposite slope. At different 
snapshots of time, i.e., $9.68$~$fs$, $27.33$~$fs$ and $32.65$~$fs$, these properties are shown in Fig.~\ref{fig:
snapshots weak}, along with the nuclear density and its components on the BO states. The notation used in
the figures is the same as in Fig.~\ref{fig: snapshots strong}.

\begin{figure}[h!]
 \begin{center}
 \includegraphics{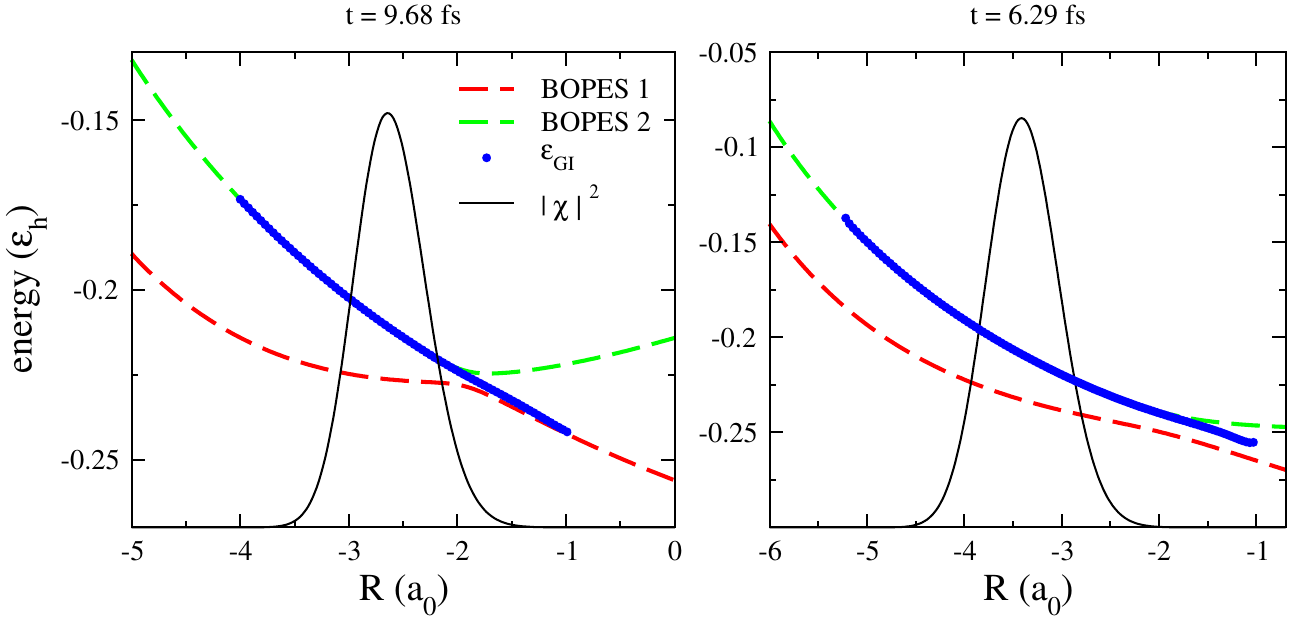}
 \end{center}
 \caption{Diabatization feature of $\epsilon_{GI}(R,t)$ (blue dots) for the two model systems (left panel, 
strong coupling at $t=9.68$~$fs$, and right panel, weak coupling at $t=6.29$~$fs$) presented here. The dashed
lines represent the BO surfaces ($\epsilon_{BO}^{(1)}(R)$ red line and $\epsilon_{BO}^{(2)}(R)$ green line) and
the continuous black line represents the nuclear density (reduced by a factor 10 and rigidly shifted along the
$y$-axis).}
 \label{fig: diabatization}
\end{figure}

A slightly different behavior from the situation of strong non-adiabatic coupling can be identified in 
$\epsilon_{GI}(R,t)$ before the passage through the avoided crossing. As the nuclear wave-packet approaches the
avoided crossing, the GI part of the TDPES ``opens'' towards the direction of motion, resembling the diabatic
surface that connects the BO surfaces through the avoided crossing. This is clearly shown in Fig.~\ref{fig:
diabatization} (left) at time $t=9.68$~$fs$ for the strongly coupled system. In the case of weaker
non-adiabatic coupling, $\epsilon_{GI}(R,t)$, at the avoided crossing, lies between the BO surfaces, as
shown in Fig.~\ref{fig: diabatization} (right).
Therefore, the diabatization feature strictly depends on the strength of the non-adiabatic coupling and, in
general, can be viewed as a transient configuration of the GI part of the TDPES before the formation of the steps.

\section{Classical dynamics on PESs}\label{sec: dynamics}
In section~\ref{sec: pes}, we have addressed some of the generic features of the TDPES that governs the nuclear
dynamics in the presence of non-adiabatic electronic transitions. As discussed before, some of these features, in 
particular the step that bridges between the two parts of the TDPES that are parallel to the BOPESs, are
reminiscent of the jumping between the BOPESs in TSH methods~\cite{TSH}. These algorithms
are based on the mixed quantum-classical treatment of the electronic and nuclear dynamics using stochastic
jumps between BO surfaces. Therefore, an ensemble of classical trajectories with different initial conditions is
needed to achieve statistically reasonable outcomes. On the other hand, the TDPES is the exact
time-dependent potential that governs the nuclear dynamics (in general together with the vector potential) and
contains the back-reaction resulting from the exact coupling to the electronic subsystem. This brings us to
investigate how the TDPES drives the classical dynamics of point-like nuclei. 

\begin{figure}[h!]
 \begin{center}
 \includegraphics{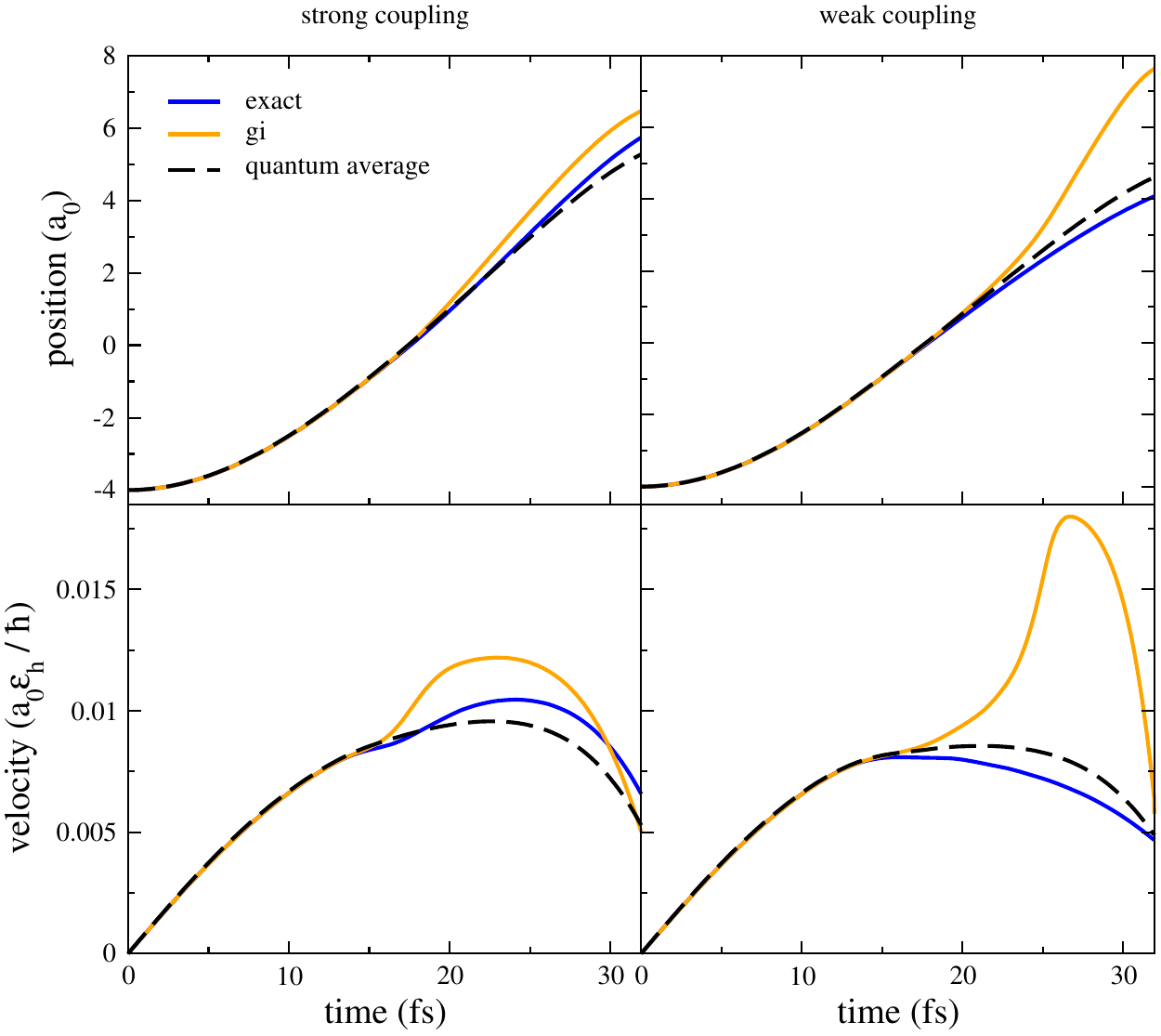}
 \end{center}
 \caption{Classical position (upper panels) and velocity (lower panels) and mean nuclear position and velocity
as functions of time for the systems in the presence of strong non-adiabatic coupling (left) and of weak
non-adiabatic coupling (right). The dashed black line represents the average nuclear values from quantum
calculation, the blue and orange lines are the positions and velocities of the classical particle when it evolves
on the exact potential and on the GI part of the potential, respectively.}
 \label{fig: position and velocity}
\end{figure}

In order to understand how the generic features of the TDPES affect the classical nuclear dynamics, we have 
employed the surfaces presented in section~(\ref{sec: pes}) to calculate the forces acting on the
nuclear degree of freedom. We compare the resulting dynamics using the forces that are calculated from the gradient
of the TDPES and from the gradient of its GI part. The classical propagation starts at the initial position
$R_c=-4.0$~$a_0$ with zero initial momentum. Here, we use the velocity-Verlet algorithm to integrate
Hamilton's equations,
\begin{equation}\label{eq: hamilton-eom}
 \left\lbrace
 \begin{array}{ccl}
  \dot R &=& \dfrac{P}{M} \\
  && \\
  \dot P &=& -\nabla_R\epsilon(R)\,\,\mbox{ or }\,\,-\nabla_R\epsilon_{GI}(R),
 \end{array}
  \right.
\end{equation}
using the same time-steps as in the quantum propagation ($\delta t = 2.4\times10^{-3}$~$fs$). In 
Fig.~\ref{fig: position and velocity} (upper panels) we present the evolution of the classical position compared
to the average nuclear position from the quantum calculation, for strong and weak coupling. In both cases, a single
trajectory, evolving on the exact surface (blue lines in Fig.~\ref{fig: position and velocity}), is able to reproduce the mean nuclear path 
(dashed black lines) fairly well. A slight deviation from the quantum results happens only towards the end of the simulated trajectories. 
When the classical forces are calculated from the GI part of the TDPES, the corresponding classical trajectory in the strong coupling case,
does not 
show a large deviation from the exact calculation. However, in the weak coupling case, after $20$~$fs$, the classical trajectory deviates
considerably 
from the quantum mean path. This behavior is also confirmed by the pronounced increase of the velocity of the classical
particle moving on $\epsilon_{GI}$, shown in Fig.~\ref{fig: position and velocity} (lower panels).

\begin{figure}[h!]
 \begin{center}
 \includegraphics[width=\textwidth]{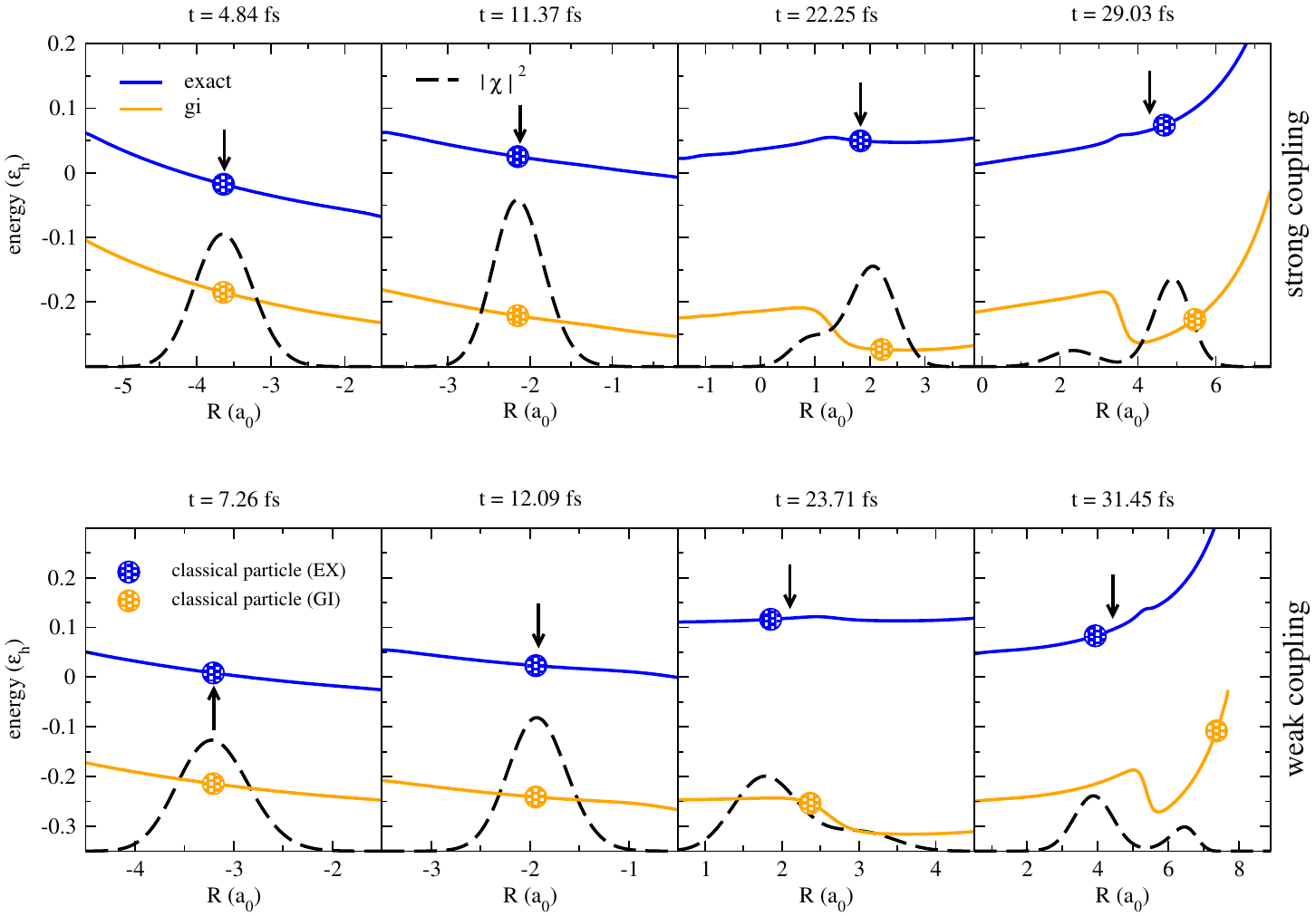}
 \end{center}
 \caption{Upper panels: strong coupling results. Lower panels: weak coupling results. The figure shows classical
positions (dots) at different times, as indicated in the plots, with the corresponding potentials,
$\epsilon_{GI}(R,t)$ (orange lines) and $\epsilon(R,t)$ (blue lines). The nuclear density (dashed black line) is
plotted as reference, along with the mean position (black arrows).}
 \label{fig: evolution}
\end{figure}

We now have a closer look at the classical dynamics and try to find out the source of the deviations, especially in
the weaker coupling case. Fig.~\ref{fig: evolution} shows the classical positions calculated from the full TDPES
(blue dots) and the GI part of it (orange dots) together with the corresponding potentials and the exact nuclear
densities at the times indicated in the plots. It can be seen in the figure that the classical particle evolving on
the GI part of the potential, in the case of weaker coupling, at the moment of the step formation feels an intense
force, as its position is exactly in the region of the step (see $t=23.71$~$fs$ in Fig.~\ref{fig: evolution}). This
happens also in the case of the strong coupling (see the blue line referring to the velocity in Fig.~\ref{fig:
position and velocity}, left plot), to a lesser extent and the velocity of the classical particle does not show
a strong peak. The evolution of the classical particle on the GI part, in the case of the strong coupling, shows
that the step forms in the direction of larger nuclear density (see plot at $t=22.25$~$fs$), hence, the classical
particle correctly follows the step and its position is approximately the mean nuclear position. However, in the
case of weaker coupling, the step forms in the direction of smaller nuclear density and the classical particle can
not move ``up the hill'' to follow the nuclear mean path, leading to a large deviation of the classical position from the quantum
mean value. The intense force felt by the classical particle drives it to an unphysical region, where the nuclear density is very small.
The presence of the GD part of the TDPES is responsible for the decrease (or even the inversion) of the ``energy gap'' in the GI part, thus
producing a better agreement between classical and quantum results. 

From comparing the classical and quantum dynamics shown in Fig.~\ref{fig: evolution}, we observe that in the strong coupling case (upper
panel), 
at $t=4.84$~$fs$ and at $t=11.37$~$fs$, the nuclear wave-packet has not yet crossed the avoided crossing, thus the GD part of the TDPES is a
constant. 
Therefore, the classical force calculated from the TDPES is identical with the one calculated from its GI part. At these times, the
classical
positions of the nuclei evolving on the GI part of the potential (orange dots in the figure) and on the full TDPES (blue dots) coincide with
the 
mean position of the nuclear wave-packet (black arrows). On the other hand, in the weaker coupling case (lower panels), a similar behavior
is seen only before the wave-packet splitting, at $t=7.26$~$fs$ and $t=12.09$~$fs$. At later times, namely $t=22.25$~$fs$ for the strong
coupling case and $t=23.71$~$fs$ for the weaker coupling case, the steps develop in $\epsilon_{GI}$ and the classical particle evolving on
this potential follows the direction in which the step is forming: in the case of strong coupling, this region coincides with the region
associated with larger nuclear density, whereas this is not the case for the weaker coupling case. As discussed above, this feature explains
why the positions of the particles on $\epsilon$ and on $\epsilon_{GI}$, for the system in the presence of strong non-adiabatic coupling,
are close to each other also at later times ($t=29.03$~$fs$ in Fig.~\ref{fig: evolution}), whereas they deviate in the weaker coupling
regime as clearly shown in the figure at time $t=31.45$~$fs$. 

The results presented in this section offer interesting insights into possible ways of modeling non-adiabatic processes, within a mixed
quantum-classical treatment. On one hand, the gradient of the GI part of the exact potential is the force that drives the classical nuclear
motion and we have shown that such force is ``adiabatic'' in the sense that, far from the step, it is produced by a single BOPES. On the
other hand, the GD part does not affect such force, but contributes in diminishing the energy separation between the two sides of the step.
This energy barrier almost disappears in the full TDPES, but the difference in slopes indeed persists. If a gauge is chosen such
that $\epsilon_{GD}(R,t)\equiv 0$, the non-zero vector potential compensates the effect of the energy step in the GI part of the TDPES by
adding a kinetic energy contribution (the vector potential appears in the kinetic term of the nuclear Hamiltonian in Eq.~(\ref{eqn: exact
nuclear eqn})). Such contribution would energetically favor the transfer of classical point-particles from one side of the step to the
other. Once again, the comparison with TSH is inevitable: in the latter, different adiabatic surfaces are energetically accessible by the
classical nuclei because of the \textsl{stochastic} jumps and the subsequent momentum rescaling (in order to impose energy conservation); in
the scheme based on the exact TDPES, depending on the gauge, either the GD part of the potential is responsible for bringing ``energetically
closer'' different BOPES or the vector potential gives the necessary kinetic energy contribution. So far, we have described where the steps
appear, how they form and how they affect nuclear motion. From these observations, we expect that rigorous mixed quantum-classical
schemes for dealing with non-adiabatic processes can be deduced in a systematic way from the classical forces associated with the exact
TDPES and the exact vector potential.

\section{Ehrenfest theorem for the nuclear wave-function}\label{sec: ehrenfest}
In section~\ref{sec: dynamics}, we studied the classical nuclear dynamics on the TDPES. However, we did not 
provide any argument on how that study can be associated with a classical limit of the nuclear motion that is able
to, approximately, reproduce the expectation values of the nuclear position and momentum of the complete
electron-nuclear system. Here, using the Ehrenfest theorem, we show how the nuclear position and momentum
calculated from Eq.~(\ref{eq: hamilton-eom}) can be linked to the expectation values of the nuclear position and
momentum of the complete electron-nuclear system. 

The Ehrenfest theorem~\cite{ehrenfest} relates the time-derivative of the expectation value of a quantum-mechanical operator $\hat
O$ to the expectation value of the commutator of that operator with the Hamiltonian, i.e.
\begin{equation}
 \frac{d}{dt}\langle\hat O(t)\rangle = \frac{1}{i\hbar}\left\langle\left[\hat O(t),\hat H\right]\right\rangle+
 \langle\partial_t\hat O(t)\rangle.
\end{equation}
The second term on the RHS refers to the explicit time-dependence of $\hat O$. In particular, the theorem leads 
to the classical-like equations of motion for the mean value of position and momentum operators. For a system of
electrons and nuclei, described by the Hamiltonian in Eq.~(\ref{eqn: hamiltonian}) and the wave-function
$\Psi(\dulr,\dulR,t)$, the mean values of the $\nu$-th nuclear position $\hat{\bf R}_{\nu}$ and momentum $\hat{\bf
P}_{\nu}$ operators evolve according to the classical Hamilton's equations
\begin{eqnarray}
 \frac{d}{dt}\langle\hat{\bf R}_{\nu}\rangle_{\Psi}=\frac{1}{i\hbar} \left\langle\left[\hat{\bf R}_\nu,
 \hat H(\dulr,\dulR)\right]\right\rangle_\Psi&=&\frac{\langle\hat{\bf P}_{\nu}\rangle_{\Psi}}{M_{\nu}}
 \label{eqn: general ehrenfest 1}\\
 \frac{d}{dt}\langle\hat{\bf P}_{\nu}\rangle_{\Psi}=\frac{1}{i\hbar} \left\langle\left[\hat{\bf P}_\nu,
 \hat H(\dulr,\dulR)\right]\right\rangle_\Psi&=& \langle-\nabla_{\nu}\big(\hat{V}_{en}(\dulr,\dulR)+\hat{W}_{nn}(\dulR)\big)
\rangle_{{\Psi}}.
 \label{eqn: general ehrenfest 2}
\end{eqnarray}
Here, the operators do not depend explicitly on time and we indicate the integration over the full
wave-function (electronic and nuclear coordinates) by $\langle\,\cdot\,\rangle_{{\Psi}}$. On the other hand, 
the nuclear equation~(\ref{eqn: exact nuclear eqn}) is a Schr\"odinger equation that contains a time-dependent
vector potential and a time-dependent scalar potential. Therefore, the Ehrenfest theorem for the nuclear subsystem
reads
\begin{eqnarray}
 \frac{d}{dt}\langle\hat{\bf R}_{\nu}\rangle_\chi&=&\frac{1}{i\hbar}\left\langle\left[\hat{\bf R}_\nu,
 \hat H_n(\dulR)\right]\right\rangle_\chi\label{eqn: ehrenfest 1}\\
 \frac{d}{dt}\langle\hat{\widetilde{\bf P}}_{\nu}\rangle_\chi&=&\frac{1}{i\hbar}\left\langle\left[
 \hat{\widetilde{\bf P}}_\nu,\hat H_n(\dulR)\right]\right\rangle_\chi+
 \left\langle\partial_t{\bf A}_\nu(\dulR,t)\right\rangle_\chi  \label{eqn: ehrenfest 2}
\end{eqnarray}
where~\cite{AMG2} 
\begin{equation}
\hat{\widetilde{\bf P}}_{\nu} = -i\hbar\nabla_\nu+{\bf A}_{\nu}(\dulR,t)
\end{equation}
is the expression of the nuclear canonical momentum operator in position representation, and 
\begin{equation}
\hat H_n(\dulR) = \sum_{\nu=1}^{N_n} \frac{\left[-i\hbar\nabla_\nu+\bA_\nu(\dulR,t)\right]^2}{2M_\nu} + 
\epsilon(\dulR,t) \label{eqn: nuclear-Hamiltonian}
\end{equation}
is the nuclear Hamiltonian from Eq.~(\ref{eqn: exact nuclear eqn}). Note that the average operation is
performed only on the nuclear wave-function as indicated by $\langle\,\cdot\,\rangle_{\chi}$. An explicit
time-dependence appears in the expression of the momentum operator, due to the presence of the vector potential.
This dependence is accounted for in the second term on the RHS of Eq.~(\ref{eqn: ehrenfest 2}). While Eq.~(\ref{eqn: ehrenfest 1}) is easily
obtained from Eq.~(\ref{eqn: general ehrenfest 1}) by
performing the integration over the electronic part of full wave-function, Eq.~(\ref{eqn: ehrenfest 2}) is more
involved and will be proved as follows. We rewrite LHS of Eq.~(\ref{eqn: general ehrenfest 2}) as
\begin{align}
 \frac{d}{dt}\langle\hat{\bf P}_{\nu}\rangle_{\Psi}=&\int d\dulr d\dulR\,
 \left[\Phi_\dulR^*(\dulr,t)\partial_t\chi^*(\dulR,t)+\chi^*(\dulR,t)\partial_t\Phi_\dulR^*(\dulr,t)\right]
 \hat{\bf P}_\nu\chi(\dulR,t)\Phi_\dulR(\dulr,t) \nonumber\\
 &+\int d\dulr d\dulR\,
 \chi^*(\dulR,t)\Phi_\dulR^*(\dulr,t)\hat{\bf P}_\nu
 \left[\Phi_\dulR(\dulr,t)\partial_t\chi(\dulR,t)+\chi(\dulR,t)\partial_t\Phi_\dulR(\dulr,t)\right].
\end{align}
$\hat{\bf P}_\nu$ being a differential operator in position representation, its action on the factorized
wave-function is
\begin{equation}
 \hat{\bf P}_\nu\chi(\dulR,t)\Phi_\dulR(\dulr,t)= \left(\hat{\bf P}_\nu\chi(\dulR,t)\right)\Phi_\dulR(\dulr,t)+
 \chi(\dulR,t)\left(\hat{\bf P}_\nu\Phi_\dulR(\dulr,t)\right).
\end{equation}
Then we use the nuclear equation~(\ref{eqn: exact nuclear eqn}) for
\begin{equation}
 \partial_t\chi(\dulR,t)=\frac{1}{i\hbar}\hat H_n(\dulR) \chi(\dulR,t)
\end{equation}
and its complex-conjugated ($\hat H_n(\dulR)$ is hermitian), the definition of the (real) vector potential
\begin{equation}
 {\bf A}_\nu(\dulR,t) = \int d\dulr \,\Phi_\dulR^*(\dulr,t) \hat{\bf P}_\nu\Phi_\dulR(\dulr,t)
\end{equation}
and the PNC, to derive
\begin{align}
 \frac{d}{dt}\langle\hat{\bf P}_{\nu}\rangle_{\Psi}=\frac{1}{i\hbar}\int d\dulR\,\chi^*(\dulR,t)
 &\left(\hat{\widetilde{\bf P}}_\nu\hat H_n(\dulR)-\hat H_n(\dulR)\hat{\widetilde{\bf P}}_\nu\right)\chi(\dulR,t)
 \nonumber \\
 +\int d\dulR\left|\chi(\dulR,t)\right|^2 \int d\dulr &\left[\left(\partial_t\Phi_\dulR^*(\dulr,t)\right)
 \hat{\bf P}_\nu\Phi_\dulR(\dulr,t) + \Phi_\dulR^*(\dulr,t)\hat{\bf P}_\nu\partial_t\Phi_\dulR(\dulr,t)\right]
\end{align}
with $\hat{\widetilde{\bf P}}_\nu=\hat{\bf P}_\nu+{\bf A}_\nu(\dulR,t)$. Using the relation
\begin{equation}
 \left(\partial_t\Phi_\dulR^*(\dulr,t)\right)\hat{\bf P}_\nu\Phi_\dulR(\dulr,t) =
 \partial_t\left(\Phi_\dulR^*(\dulr,t)\hat{\bf P}_\nu\Phi_\dulR(\dulr,t)\right) -
 \Phi_\dulR^*(\dulr,t)\hat{\bf P}_\nu\partial_t\Phi_\dulR(\dulr,t),
\end{equation}
for the term in the square brackets, leads to
\begin{align}
 \frac{d}{dt}\langle\hat{\bf P}_{\nu}\rangle_{\Psi}=\int d\dulR\,\chi^*(\dulR,t)\left(
 \frac{1}{i\hbar}\left[\hat{\widetilde{\bf P}}_\nu,\hat H_n(\dulR)\right]+\partial_t{\bf A}_\nu(\dulR,t)\right)
 \chi(\dulR,t),
\end{align}
recovering the term on the RHS of Eq.~(\ref{eqn: ehrenfest 2}). A similar procedure \cite{AMG2} yields the relation
\begin{align}
 \langle\hat{\bf P}_\nu\rangle_\Psi &= \int d\dulr d\dulR \,\Phi_\dulR^*(\dulr,t)\chi^*(\dulR,t)
 \left[\left(\hat{\bf P}_\nu\chi(\dulR,t)\right)\Phi_\dulR(\dulr,t)+\chi(\dulR,t)\hat{\bf
 P}_\nu\Phi_\dulR(\dulr,t)\right] \nonumber \\
 &=\int d\dulR \,\chi^*(\dulR,t)\left[\hat{\bf P}_\nu+{\bf A}_\nu(\dulR,t)\right]\chi(\dulR,t)=\langle\hat{\widetilde{\bf
P}}_\nu\rangle_\chi,
\end{align}
which proves the identity of the LHSs of Eqs.~(\ref{eqn: general ehrenfest 2}) and~(\ref{eqn: ehrenfest 2}).

We have proved the Ehrenfest theorem for the nuclear wave-function and nuclear Hamiltonian, deriving exact relations for the evolution of
the mean values of nuclear position and momentum operators over the complete system. This outcome is consistent with the interpretation of
$\chi(\dulR,t)$ as the proper nuclear wave-function that reproduces the nuclear density and current density of the complete system (see
the discussion in section~\ref{sec: background}).

In the one-dimensional system studied here, the gauge is chosen such that $A(R,t)=0$, therefore, the Ehrenfest equations become
\begin{eqnarray}
 \frac{d}{dt}\langle\hat R\rangle_\chi=\frac{1}{i\hbar}\left\langle\left[\hat R,\hat H_n\right]\right\rangle_\chi
 &=&\frac{\langle\hat P\rangle_\chi}{M}\label{eqn: ehrenfest 1 one-d}\\
 \frac{d}{dt}\langle\hat P\rangle_\chi=\frac{1}{i\hbar}\left\langle\left[\hat P,\hat H_n\right]\right\rangle_\chi
 &=&\langle-\nabla_R\epsilon(R,t)\rangle_\chi, \label{eqn: ehrenfest 2 one-d}
\end{eqnarray}
where the mean force generating the classical-like evolution is determined as the expectation value, on the nuclear wave-function, of the
gradient of the TDPES. If we replace the nuclear wave-function in Eqs.~(\ref{eqn: ehrenfest 1 one-d}) and~(\ref{eqn: ehrenfest 2 one-d}) by
a delta-function centered at the classical position, we get Eqs.~(\ref{eq: hamilton-eom}) that was used in section~\ref{sec: dynamics}
to generate classical dynamics on the exact PES. That is why the classical nuclear dynamics on the TDPES could actually approximate the mean
nuclear position and momentum.

We have numerically simulated classical dynamics under the following equations of motion
\begin{equation}
 \left\lbrace
 \begin{array}{ccl}
 \dot R &=&\dfrac{P}{M}\\
  && \\
 \dot P&=&\langle-\nabla_R\epsilon(R,t)\rangle_\chi,
 \end{array}
  \right.
\end{equation}
where $\epsilon(R,t)$ is obtained from the solution of the TDSE with Hamiltonian~(\ref{eqn:
metiu-hamiltonian}), for both sets of parameters producing strong and weak non-adiabatic coupling between the two
lowest BO surfaces. The initial conditions for the classical evolution are exactly the initial mean position and
mean velocity of the quantum particle.
\begin{figure}[h!]
 \begin{center}
 \includegraphics{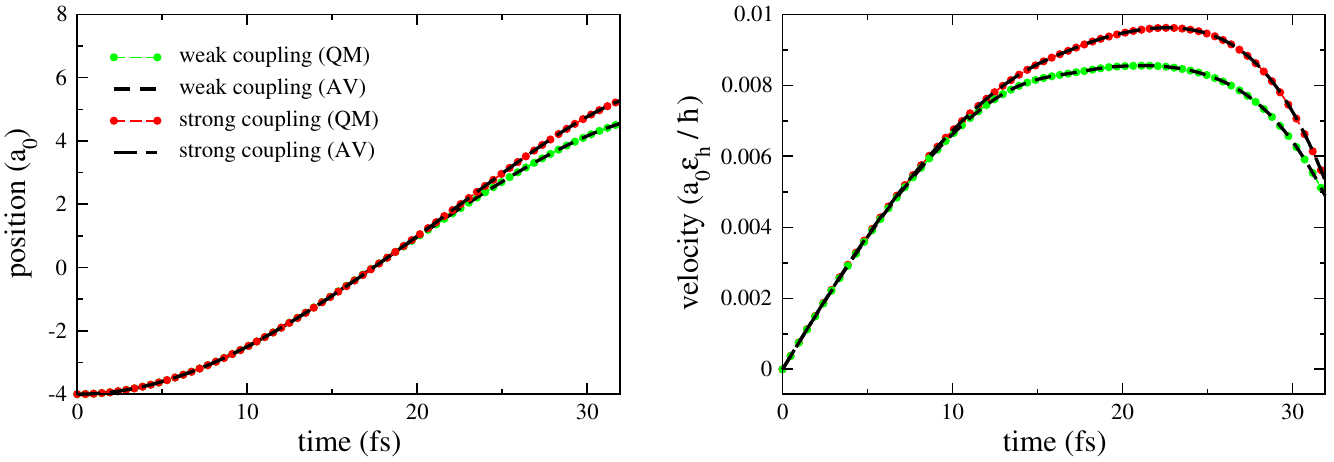}
 \end{center}
 \caption{Left: nuclear position as a function of time. Right: nuclear velocity as a function of time. The
average position and velocity calculated from the quantum-mechanical (QM) propagation are shown as dotted red
(strong coupling) and dotted green (weak coupling) lines. The long-dashed (strong coupling) and short-dashed (weak
coupling) black lines are the results of classical propagation driven by the average force (AV) as in
Eqs.~(\ref{eqn: ehrenfest 1 one-d}) and~(\ref{eqn: ehrenfest 2 one-d}).}
 \label{fig: ehrenfest}
\end{figure}
The results are shown in Fig.~\ref{fig: ehrenfest}, where we plot the mean position (left) and velocity
(right) as functions of time from quantum-mechanical calculations, compared to the values of position and velocity
of a classical particle moving according to the average force $\langle-\nabla_R\epsilon(R,t)\rangle_\chi$. As
expected by the proof of the Ehrenfest theorem involving the nuclear wave-function
$\chi(\dulR,t)$ and the nuclear Hamiltonian $\hat H_n$ presented in this section, the classical trajectory perfectly
follows the evolution of the quantum mean values.
In section~\ref{sec: dynamics}, we studied the classical nuclear dynamics on the TDPES. However, we did not 
provide any argument on how that study can be associated with a classical limit of the nuclear motion that is able
to, approximately, reproduce the expectation values of the nuclear position and momentum of the complete
electron-nuclear system. Here, using the Ehrenfest theorem, we show how the nuclear position and momentum
calculated from Eq.~(\ref{eq: hamilton-eom}) can be linked to the expectation values of the nuclear position and
momentum of the complete electron-nuclear system. 

The Ehrenfest theorem~\cite{ehrenfest} relates the time-derivative of the expectation value of a quantum-mechanical operator $\hat
O$ to the expectation value of the commutator of that operator with the Hamiltonian, i.e.
\begin{equation}
 \frac{d}{dt}\langle\hat O(t)\rangle = \frac{1}{i\hbar}\left\langle\left[\hat O(t),\hat H\right]\right\rangle+
 \langle\partial_t\hat O(t)\rangle.
\end{equation}
The second term on the RHS refers to the explicit time-dependence of $\hat O$. In particular, the theorem leads 
to the classical-like equations of motion for the mean value of position and momentum operators. For a system of
electrons and nuclei, described by the Hamiltonian in Eq.~(\ref{eqn: hamiltonian}) and the wave-function
$\Psi(\dulr,\dulR,t)$, the mean values of the $\nu$-th nuclear position $\hat{\bf R}_{\nu}$ and momentum $\hat{\bf
P}_{\nu}$ operators evolve according to the classical Hamilton's equations
\begin{eqnarray}
 \frac{d}{dt}\langle\hat{\bf R}_{\nu}\rangle_{\Psi}=\frac{1}{i\hbar} \left\langle\left[\hat{\bf R}_\nu,
 \hat H(\dulr,\dulR)\right]\right\rangle_\Psi&=&\frac{\langle\hat{\bf P}_{\nu}\rangle_{\Psi}}{M_{\nu}}
 \label{eqn: general ehrenfest 1}\\
 \frac{d}{dt}\langle\hat{\bf P}_{\nu}\rangle_{\Psi}=\frac{1}{i\hbar} \left\langle\left[\hat{\bf P}_\nu,
 \hat H(\dulr,\dulR)\right]\right\rangle_\Psi&=& \langle-\nabla_{\nu}\big(\hat{V}_{en}(\dulr,\dulR)+\hat{W}_{nn}(\dulR)\big)
\rangle_{{\Psi}}.
 \label{eqn: general ehrenfest 2}
\end{eqnarray}
Here, the operators do not depend explicitly on time and we indicate the integration over the full
wave-function (electronic and nuclear coordinates) by $\langle\,\cdot\,\rangle_{{\Psi}}$. On the other hand, 
the nuclear equation~(\ref{eqn: exact nuclear eqn}) is a Schr\"odinger equation that contains a time-dependent
vector potential and a time-dependent scalar potential. Therefore, the Ehrenfest theorem for the nuclear subsystem
reads
\begin{eqnarray}
 \frac{d}{dt}\langle\hat{\bf R}_{\nu}\rangle_\chi&=&\frac{1}{i\hbar}\left\langle\left[\hat{\bf R}_\nu,
 \hat H_n(\dulR)\right]\right\rangle_\chi\label{eqn: ehrenfest 1}\\
 \frac{d}{dt}\langle\hat{\widetilde{\bf P}}_{\nu}\rangle_\chi&=&\frac{1}{i\hbar}\left\langle\left[
 \hat{\widetilde{\bf P}}_\nu,\hat H_n(\dulR)\right]\right\rangle_\chi+
 \left\langle\partial_t{\bf A}_\nu(\dulR,t)\right\rangle_\chi  \label{eqn: ehrenfest 2}
\end{eqnarray}
where~\cite{AMG2} 
\begin{equation}
\hat{\widetilde{\bf P}}_{\nu} = -i\hbar\nabla_\nu+{\bf A}_{\nu}(\dulR,t)
\end{equation}
is the expression of the nuclear canonical momentum operator in position representation, and 
\begin{equation}
\hat H_n(\dulR) = \sum_{\nu=1}^{N_n} \frac{\left[-i\hbar\nabla_\nu+\bA_\nu(\dulR,t)\right]^2}{2M_\nu} + 
\epsilon(\dulR,t) \label{eqn: nuclear-Hamiltonian}
\end{equation}
is the nuclear Hamiltonian from Eq.~(\ref{eqn: exact nuclear eqn}). Note that the average operation is
performed only on the nuclear wave-function as indicated by $\langle\,\cdot\,\rangle_{\chi}$. An explicit
time-dependence appears in the expression of the momentum operator, due to the presence of the vector potential.
This dependence is accounted for in the second term on the RHS of Eq.~(\ref{eqn: ehrenfest 2}). While Eq.~(\ref{eqn: ehrenfest 1}) is easily
obtained from Eq.~(\ref{eqn: general ehrenfest 1}) by
performing the integration over the electronic part of full wave-function, Eq.~(\ref{eqn: ehrenfest 2}) is more
involved and will be proved as follows. We rewrite LHS of Eq.~(\ref{eqn: general ehrenfest 2}) as
\begin{align}
 \frac{d}{dt}\langle\hat{\bf P}_{\nu}\rangle_{\Psi}=&\int d\dulr d\dulR\,
 \left[\Phi_\dulR^*(\dulr,t)\partial_t\chi^*(\dulR,t)+\chi^*(\dulR,t)\partial_t\Phi_\dulR^*(\dulr,t)\right]
 \hat{\bf P}_\nu\chi(\dulR,t)\Phi_\dulR(\dulr,t) \nonumber\\
 &+\int d\dulr d\dulR\,
 \chi^*(\dulR,t)\Phi_\dulR^*(\dulr,t)\hat{\bf P}_\nu
 \left[\Phi_\dulR(\dulr,t)\partial_t\chi(\dulR,t)+\chi(\dulR,t)\partial_t\Phi_\dulR(\dulr,t)\right].
\end{align}
$\hat{\bf P}_\nu$ being a differential operator in position representation, its action on the factorized
wave-function is
\begin{equation}
 \hat{\bf P}_\nu\chi(\dulR,t)\Phi_\dulR(\dulr,t)= \left(\hat{\bf P}_\nu\chi(\dulR,t)\right)\Phi_\dulR(\dulr,t)+
 \chi(\dulR,t)\left(\hat{\bf P}_\nu\Phi_\dulR(\dulr,t)\right).
\end{equation}
Then we use the nuclear equation~(\ref{eqn: exact nuclear eqn}) for
\begin{equation}
 \partial_t\chi(\dulR,t)=\frac{1}{i\hbar}\hat H_n(\dulR) \chi(\dulR,t)
\end{equation}
and its complex-conjugated ($\hat H_n(\dulR)$ is hermitian), the definition of the (real) vector potential
\begin{equation}
 {\bf A}_\nu(\dulR,t) = \int d\dulr \,\Phi_\dulR^*(\dulr,t) \hat{\bf P}_\nu\Phi_\dulR(\dulr,t)
\end{equation}
and the PNC, to derive
\begin{align}
 \frac{d}{dt}\langle\hat{\bf P}_{\nu}\rangle_{\Psi}=\frac{1}{i\hbar}\int d\dulR\,\chi^*(\dulR,t)
 &\left(\hat{\widetilde{\bf P}}_\nu\hat H_n(\dulR)-\hat H_n(\dulR)\hat{\widetilde{\bf P}}_\nu\right)\chi(\dulR,t)
 \nonumber \\
 +\int d\dulR\left|\chi(\dulR,t)\right|^2 \int d\dulr &\left[\left(\partial_t\Phi_\dulR^*(\dulr,t)\right)
 \hat{\bf P}_\nu\Phi_\dulR(\dulr,t) + \Phi_\dulR^*(\dulr,t)\hat{\bf P}_\nu\partial_t\Phi_\dulR(\dulr,t)\right]
\end{align}
with $\hat{\widetilde{\bf P}}_\nu=\hat{\bf P}_\nu+{\bf A}_\nu(\dulR,t)$. Using the relation
\begin{equation}
 \left(\partial_t\Phi_\dulR^*(\dulr,t)\right)\hat{\bf P}_\nu\Phi_\dulR(\dulr,t) =
 \partial_t\left(\Phi_\dulR^*(\dulr,t)\hat{\bf P}_\nu\Phi_\dulR(\dulr,t)\right) -
 \Phi_\dulR^*(\dulr,t)\hat{\bf P}_\nu\partial_t\Phi_\dulR(\dulr,t),
\end{equation}
for the term in the square brackets, leads to
\begin{align}
 \frac{d}{dt}\langle\hat{\bf P}_{\nu}\rangle_{\Psi}=\int d\dulR\,\chi^*(\dulR,t)\left(
 \frac{1}{i\hbar}\left[\hat{\widetilde{\bf P}}_\nu,\hat H_n(\dulR)\right]+\partial_t{\bf A}_\nu(\dulR,t)\right)
 \chi(\dulR,t),
\end{align}
recovering the term on the RHS of Eq.~(\ref{eqn: ehrenfest 2}). A similar procedure \cite{AMG2} yields the relation
\begin{align}
 \langle\hat{\bf P}_\nu\rangle_\Psi &= \int d\dulr d\dulR \,\Phi_\dulR^*(\dulr,t)\chi^*(\dulR,t)
 \left[\left(\hat{\bf P}_\nu\chi(\dulR,t)\right)\Phi_\dulR(\dulr,t)+\chi(\dulR,t)\hat{\bf
 P}_\nu\Phi_\dulR(\dulr,t)\right] \nonumber \\
 &=\int d\dulR \,\chi^*(\dulR,t)\left[\hat{\bf P}_\nu+{\bf A}_\nu(\dulR,t)\right]\chi(\dulR,t)=\langle\hat{\widetilde{\bf
P}}_\nu\rangle_\chi,
\end{align}
which proves the identity of the LHSs of Eqs.~(\ref{eqn: general ehrenfest 2}) and~(\ref{eqn: ehrenfest 2}).

We have proved the Ehrenfest theorem for the nuclear wave-function and nuclear Hamiltonian, deriving exact relations for the evolution of
the mean values of nuclear position and momentum operators over the complete system. This outcome is consistent with the interpretation of
$\chi(\dulR,t)$ as the proper nuclear wave-function that reproduces the nuclear density and current density of the complete system (see
the discussion in section~\ref{sec: background}).

In the one-dimensional system studied here, the gauge is chosen such that $A(R,t)=0$, therefore, the Ehrenfest equations become
\begin{eqnarray}
 \frac{d}{dt}\langle\hat R\rangle_\chi=\frac{1}{i\hbar}\left\langle\left[\hat R,\hat H_n\right]\right\rangle_\chi
 &=&\frac{\langle\hat P\rangle_\chi}{M}\label{eqn: ehrenfest 1 one-d}\\
 \frac{d}{dt}\langle\hat P\rangle_\chi=\frac{1}{i\hbar}\left\langle\left[\hat P,\hat H_n\right]\right\rangle_\chi
 &=&\langle-\nabla_R\epsilon(R,t)\rangle_\chi, \label{eqn: ehrenfest 2 one-d}
\end{eqnarray}
where the mean force generating the classical-like evolution is determined as the expectation value, on the nuclear wave-function, of the
gradient of the TDPES. If we replace the nuclear wave-function in Eqs.~(\ref{eqn: ehrenfest 1 one-d}) and~(\ref{eqn: ehrenfest 2 one-d}) by
a delta-function centered at the classical position, we get Eqs.~(\ref{eq: hamilton-eom}) that was used in section~\ref{sec: dynamics}
to generate classical dynamics on the exact PES. That is why the classical nuclear dynamics on the TDPES could actually approximate the mean
nuclear position and momentum.

We have numerically simulated classical dynamics under the following equations of motion
\begin{equation}
 \left\lbrace
 \begin{array}{ccl}
 \dot R &=&\dfrac{P}{M}\\
  && \\
 \dot P&=&\langle-\nabla_R\epsilon(R,t)\rangle_\chi,
 \end{array}
  \right.
\end{equation}
where $\epsilon(R,t)$ is obtained from the solution of the TDSE with Hamiltonian~(\ref{eqn:
metiu-hamiltonian}), for both sets of parameters producing strong and weak non-adiabatic coupling between the two
lowest BO surfaces. The initial conditions for the classical evolution are exactly the initial mean position and
mean velocity of the quantum particle.
\begin{figure}[h!]
 \begin{center}
 \includegraphics{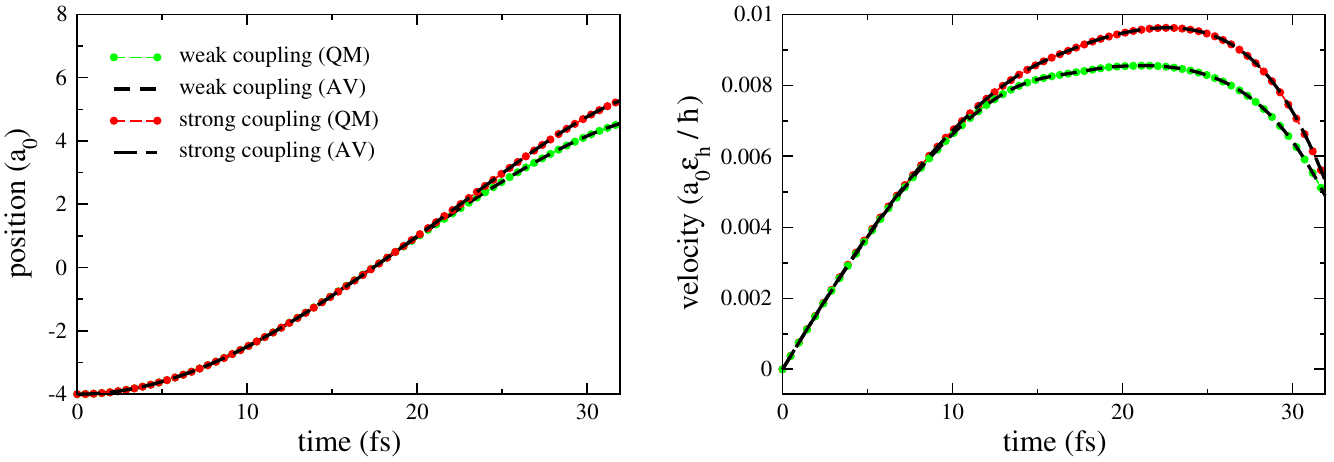}
 \end{center}
 \caption{Left: nuclear position as a function of time. Right: nuclear velocity as a function of time. The
average position and velocity calculated from the quantum-mechanical (QM) propagation are shown as dotted red
(strong coupling) and dotted green (weak coupling) lines. The long-dashed (strong coupling) and short-dashed (weak
coupling) black lines are the results of classical propagation driven by the average force (AV) as in
Eqs.~(\ref{eqn: ehrenfest 1 one-d}) and~(\ref{eqn: ehrenfest 2 one-d}).}
 \label{fig: ehrenfest}
\end{figure}
The results are shown in Fig.~\ref{fig: ehrenfest}, where we plot the mean position (left) and velocity
(right) as functions of time from quantum-mechanical calculations, compared to the values of position and velocity
of a classical particle moving according to the average force $\langle-\nabla_R\epsilon(R,t)\rangle_\chi$. As
expected by the proof of the Ehrenfest theorem involving the nuclear wave-function
$\chi(\dulR,t)$ and the nuclear Hamiltonian $\hat H_n$ presented in this section, the classical trajectory perfectly
follows the evolution of the quantum mean values.

\section{Conclusion}\label{sec: conclusion}
In a system of interacting electrons and nuclei, the nuclear dynamics is fully determined by the TDPES and the 
time-dependent vector potential defined in the framework of the exact decomposition of the electronic and nuclear 
motions, as presented in this paper. We investigated some situations in which the vector potential can be gauged
away, thus making the TDPES responsible for the nuclear evolution. This time-dependent scalar potential presents
distinct and general features that can be analyzed in terms of its GI and GD components. The former, (i) in the
region of an avoided crossing has a pronounced \textsl{diabatic} character, smoothly connecting different BOPESs
along the direction of the nuclear wave-packet's motion, and, (ii) further away from the avoided crossing, \textsl{dynamical
steps} appear between regions in which the (GI part of the) exact potential coincides with one or the other BOPES.
The latter is either constant, if the nuclear wave-packet does not split, or stepwise constant, with the step at
the same position, and with opposite slope, as in the GI part of the TDPES. We have analyzed in detail these
features and discussed the connections with a classical picture of the nuclear evolution. To this end, we
calculated the classical forces from the TDPES and from its GI component and performed classical nuclear dynamics
driven by those forces. The importance of the GD part of the potential is evident as it improves the agreement of
classical results with the quantum-mechanical calculations. We conclude that, if the exact TDPES is available, a
single classical trajectory is able to reproduce quantum results fairly well, as long as quantum nuclear effects,
such as tunneling or splitting of the nuclear wave-packet, are negligible. We have seen, in the example presented
in the paper, that the splitting of the nuclear wave-function at the avoided crossing, that cannot be captured in
the classical study, is responsible for the deviation of the classical results from the expected quantum behavior.
Further analysis involving the propagation of multiple independent trajectories on the exact TDPES are envisaged.
Such a multi-trajectory approach should be able to reproduce non-adiabatic effects, as those described above.

The development of mixed quantum-classical schemes to treat the non-adiabatic coupled electron-nuclear 
dynamics is still a challenging topic in physics and chemistry. Investigating the properties of the exact
potential, that incorporates the effects of the electronic quantum dynamics on the nuclei, is a first step towards
understanding the key features of approximated potentials and algorithms. We did not consider here cases where the
vector potential cannot be gauged away. This will be the subject of future investigations. 

In the final part of the paper, we have shown that the Ehrenfest theorem applied to calculate the mean nuclear 
position and momentum based on the nuclear equation alone reproduces the mean values calculated from the complete
electron-nuclear system.

\section*{Acknowledgements}
Partial support from the Deutsche Forschungsgemeinschaft (SFB 762) and from the European Commission (FP7-NMP-CRONOS) is gratefully
acknowledged.

\addcontentsline{toc}{section}{References}
%\bibliography{./steps}

\end{document}